\documentclass[A4paper,11pt]{article}
\usepackage{fullpage}

\usepackage{latexsym}
\usepackage{url}
\usepackage{xspace}
\usepackage{color}
\usepackage[font=small]{caption}

\usepackage{xspace}

\usepackage{cite}

\usepackage{tikz}
\usetikzlibrary{arrows.meta}
\tikzset{>={Classical TikZ Rightarrow[width=1.7mm,length=1mm]}}

\graphicspath{{./graphics/benchmarks/}}

\definecolor{mygreen}{rgb}{0,0.5,0}

\newcommand{\eg}{e.g., }
\newcommand{\ie}{i.e., }

\newcommand{\stamptrue}{\textsf{true}\xspace}
\newcommand{\stampfalse}{\textsf{false}\xspace}

\title{A new and five older Concurrent Memory Reclamation Schemes in
  Comparison (\emph{Stamp-it})}

\author{
Manuel P\"oter \\
TU Wien, Faculty of Informatics\\
Vienna, Austria\\
\url{manuel@manuel-poeter.at}
\and 
Jesper Larsson Tr\"aff\\
TU Wien, Faculty of Informatics\\
Vienna, Austria\\
\url{traff@par.tuwien.ac.at}
}

\begin{document}

\maketitle

\begin{abstract}
Memory management is a critical component in almost all shared-memory,
concurrent data structures and algorithms, consisting in the efficient
allocation and the subsequent reclamation of shared memory
resources. This paper contributes a new, lock-free, amortized
constant-time memory reclamation scheme called \emph{Stamp-it}, and
compares it to five well-known, selectively efficient schemes from the
literature, namely Lock-free Reference Counting, Hazard Pointers,
Quiescent State-based Reclamation, Epoch-based Reclamation, and New
Epoch-based Reclamation. An extensive, experimental evaluation with
both new and commonly used benchmarks is provided, on four different
shared-memory systems with hardware supported thread counts ranging
from 48 to 512, showing Stamp-it to be competitive with and in many
cases and aspects outperforming other schemes.
\end{abstract}

\section{Introduction}

Efficient, dynamic memory management is at the heart of many
sequential and parallel algorithms, and consist in the allocation of
pieces of memory and the subsequent, \emph{safe} reclamation of these
pieces when they are no longer in use.  In parallel and concurrent,
lock- and wait-free algorithms, the reclamation step is highly
non-trivial since more than one \emph{thread} may be referencing and
using an allocated piece of memory unbeknownst to other threads: It
cannot be given back to the system or thread-local heap before it has
been ascertained that no threads will possibly access any data in this
memory anymore.

Naturally, there has been a substantial amount of work on memory
reclamation for concurrent algorithms, see,
\eg~\cite{Alistarh:2014:StackTrack,Alistarh:2015:TAS:2755573.2755600,Balmau:2016:FRM:2935764.2935790,Braginsky:2013:Drop,Brown:2015:RML:2767386.2767436,Cohen:2015:AMR:2814270.2814298,Cohen:2015:EMM:2755573.2755579,Gidenstam:2009,Herlihy:2002:ROP,RamalheteC17,Michael:HazardPointers,Hart:2007:PMR:1316099.1316427,Detlefs01lock-freereference,Sundell:2005}.
All of these schemes have their merits and (performance) issues. One
drawback shared by them all, except for reference counting schemes, is
that they need to scan references from \emph{all threads} in order to
reclaim possibly no longer referenced memory pieces. A main motivation
of this work is to overcome this bound.

The contribution of this paper is a new lock-free reclamation scheme,
called \emph{Stamp-it}, which is compared qualitatively and
experimentally to five well-known and, depending on
circumstances, well performing current schemes. Reclamation in Stamp-it is
done in amortized constant time per reclaimed memory block; no
references are scanned unless they can be reclaimed.  All tested
schemes have been (re)implemented in C++; the full source code is
available at \url{http://github.com/mpoeter/emr}.  The experimental
evaluation is done on four architecturally different systems with
large numbers of hardware supported threads, ranging from 48 up to
512. We use standard benchmarks, as well as a new benchmark designed
to study memory consumption by reclaimable but unreclaimed memory.
On these benchmarks and machines Stamp-it compares favorably and in
many cases and aspects significantly outperforms the competing
schemes.

A contiguous piece of memory allocated from the system heap for use in
the concurrent algorithm and possibly shared between the threads is
called a \emph{node}. Efficient allocation and deallocation is a
complex topic on its own and a number of scalable memory managers have
been published \cite{Berger:2000:HSM:384264.379232, tcmalloc,
  Evans06ascalable, Michael:2004:SLD:996893.996848}, but is outside
the scope of this work. We mainly use the allocator from the C++
standard library; except on Solaris where we use \texttt{jemalloc} as
described later.

Nodes may store additional meta-information that is normally not seen
by the application; additional meta-information needed by the
different reclamation schemes will be discussed. We use $p$ to denote
the number of threads.

A \emph{general purpose} reclamation scheme allows the eventually
reclaimed memory of nodes to be freely reused at a later time,
regardless of how and in which data structure the allocated nodes were
used. Not all reclamation schemes have this property,
\eg~\cite{Valois:Phd,Sundell:2005} do not allow general reuse of
reclaimed nodes, \cite{Braginsky:2013:Drop,Gidenstam:2009} have to be
tailored to each data structure
and~\cite{Cohen:2015:EMM:2755573.2755579,
  Cohen:2015:AMR:2814270.2814298} require the data structure to be in
a special, normalized form.  A general purpose scheme should be
non-intrusive, requiring no or little changes in the code. A way of
achieving this is to rely on a standard interface as those proposed
for C++~\cite{Robison:2013}. A reclamation scheme should be fast, both
in use and maintenance of references to shared nodes, as well as in
the actual reclamation. It should require little memory overhead,
avoid typical performance issues like false sharing and should not
prevent data structures from using typical patterns found in lock-free
programming like borrowing some bits from a pointer.  Reclaimability
of nodes should be detected fast to reduce memory pressure.
Robustness against crashes, and bounds on the amount of memory blocked
by crashed threads are desirable.  Lock-freedom should allow for good
scalability; wait-freedom would be desirable, but not many schemes
actually provide this.

All lock- and wait-free algorithms rely on hardware supported atomic
operations. We consider only solutions that use hardware atomics
available in standard processors like \emph{fetch-and-add} (FAA) and
\emph{single-word compare-and-swap} (CAS). Solutions requiring
non-standard double-word compare-and-swap (as in,
\eg~\cite{Detlefs01lock-freereference}) will either be non-portable or
require expensive emulations. We also ruled out solutions that have to
be tailored to specific data structures
(like~\cite{Braginsky:2013:Drop, Gidenstam:2009}) or that require
hardware or operating system specific features like transactional
memory (\eg~\cite{Alistarh:2014:StackTrack}) or POSIX signals
(\eg~\cite{Alistarh:2015:TAS:2755573.2755600,Brown:2015:RML:2767386.2767436}).
The aim was to create a portable, fully C++ standard conform and
platform independent implementation. Our implementation
is mature beyond a simple proof of concept, and is applicable for real-life
applications and works with arbitrary numbers of threads that can be
started and stopped arbitrarily.

Based on the above discussion, we have implemented Lock-free Reference
Counting (LFRC)~\cite{Valois:Phd}, Hazard Pointers
(HP)~\cite{Michael:HazardPointers}, Quiescent State-based Reclamation
(QSR), Epoch-based Reclamation (ER)~\cite{Fraser:Phd}, and New
Epoch-based Reclamation
(NER)~\cite{Hart:2007:PMR:1316099.1316427}. Hart et
al. ~\cite{Hart:2007:PMR:1316099.1316427} used the same selection of
schemes in their study, and we wanted to repeat their experiments with
our own implementations on different platforms and at a larger scale.

\section{Memory Reclamation Schemes}

We first describe in more detail five memory reclamation schemes,
selected according to the desired criteria described above. This
provides the basis for introducing \emph{Stamp-it}, and qualitatively
compare it to the other schemes. Although fitting our criteria, we
have at this point in time not considered the scheme
DEBRA~\cite{Brown:2015:RML:2767386.2767436}, but may add results from
this at a later time. Another very recent scheme called ``Interval
based reclamation'' (paper to appear) is likewise not discussed.

\subsection{Lock-free Reference Counting}
\label{sec:lfrc}
Reference counting is a well known concept that has been used for
decades.  The first reclamation scheme for lock-free data structures
based on reference counting was presented by John
D.~Valois~\cite{Valois:Phd}. The original proposal contained race
conditions that were discovered and corrected by Maged M.~Michael and
Michael L.~Scott~\cite{Michael:1995:CMM:898203}.

In reference counting, each node is equipped with an integer field that
tracks the number of references to the node. Each thread is
responsible for updating this reference counter
accordingly---incrementing it for each new reference, decrementing it
for every dropped reference. The increment is implemented using a
simple FAA operation. The decrement, however, is more complicated and
has to be implemented using a CAS operation; the reason for this is
explained below.  When a reference counter drops to zero there are
no more references to this node and it can therefore be reclaimed. A
general problem with reference counting schemes is that they can only
reclaim nodes in acyclic structures; circular structures are
vulnerable to memory leakage.

Although lock-free reference counting (LFRC) avoids locks, it cannot
guarantee an upper bound on the amount of memory consumed by removed
nodes, since every thread can hold an arbitrary number of references
to nodes. It has been shown by Michael~\cite{Michael:HazardPointers}
and Hart et al.~\cite{Hart:2007:PMR:1316099.1316427} that reference
counting can incur large performance overheads that often makes
lock-free data structures perform worse than their lock-based counterparts.

LFRC is not a general reclamation scheme where reclaimed memory can be
reused arbitrarily. However, LFRC can be used in situations where
reclaimed nodes are reused in the same data structure. As the
following example shows, the scheme expects a node's reference counter
to be available indefinitely, such that it is possible to update the
reference count on a potentially reclaimed node without corrupting the
data structure. This would not be the case if the memory was reused in
different data structures.

Assume a thread wants to move from one node to the next in a linked
list. With LFRC, it has has to perform the following steps:
\begin{enumerate}
\item \label{LFRC-step-1} Read the reference to the next node.
\item \label{LFRC-step-2} Increase the reference counter of the next node.
\item \label{LFRC-step-3} Reread the reference to the next node and check
  whether it has changed in the meantime.
  \begin{enumerate}
  \item If it has changed, decrease the reference counter of the next
    node, drop the reference and start again at Step~\ref{LFRC-step-1}.
  \item Otherwise the thread has a safe reference to the next node.
  \end{enumerate}
\end{enumerate}

In this sequence there is a race condition between
Step~\ref{LFRC-step-1} (reading the reference) and
Step~\ref{LFRC-step-2} (incrementing the reference counter).  It could
happen that between these two steps the node's reference counter drops
to zero (due to another thread releasing its reference) such that the
node becomes reclaimable. In Step~\ref{LFRC-step-2}, the thread would
then increment the reference counter of a potentially reclaimed node.
To overcome the race, Step~\ref{LFRC-step-3} rereads the reference to
the next node to ensure that it has not changed in the meantime. In
case it has changed the thread has to restart the whole procedure.

In order to reuse reclaimed nodes in the same data structure, a
special, global free-list is maintained. When a thread wants to
decrement the reference count it checks whether it is about to
drop the last reference and if that is the case tries to set a
``claim bit'' 
in a single
atomic compare-and-swap operation. The thread that successfully sets
the claim bit can safely push the node to the free-list.

There are several other proposals for systems based on this reference
counting scheme. Detlefs et al.~\cite{Detlefs01lock-freereference}
allow changing the node's type upon reclamation but require a
double-compare-and-swap (DCAS) operation
which is usually not supported by current CPUs. Another scheme
proposed by Sundell~\cite{Sundell:2005} is wait-free, but the number of
threads has to be known in advance. 

\subsection{Hazard pointers}
Hazard Pointer based reclamation (HP) was introduced by
M. Mi\-chael~\cite{Michael:HazardPointers}. This scheme is sometimes
also referred to as \emph{safe memory reclamation}
(SMR)~\cite{Fraser:Phd,Michael:2002:HPD:564870.564881,SundellTsigas:2008}.
HP is based on the observation that, in the vast majority of lock-free
data structures, threads hold only a small number of references that
may later be used without further validation. 
The main idea is to associate a number of single-writer, multi-reader
shared pointers, called \emph{hazard pointers}, with each thread to
operate on the associated nodes. Each thread has $k$ hazard pointers
(depending on the actual algorithm and data structure, but $k$ must be
known \emph{a priori}). With $p$ threads, $H=pk$ hazard pointers are
needed in total.

When a thread wants to access a shared node, it stores the node's
reference in one of its unused hazard pointers. This is the way to signal
to the other threads that this thread is using this particular node and
that it is therefore not safe to reclaim it. When the thread no longer
needs the node it simply resets the according hazard pointer to null.

Nodes that have been removed from the data structure and 
need to be reclaimed (called \emph{retired nodes}
in~\cite{Michael:HazardPointers}) are maintained in thread-local
lists. Whenever the size of a thread's list reaches some chosen
threshold $R$, the thread tries to reclaim the nodes from the
list. Increasing $R$ amortizes reclamation overhead across more nodes,
but increases memory usage; if $R$ is larger than $H$ by some amount
proportional to $H$
the amortized per-node processing time is constant, but this can cause
performance issues with large numbers of threads as will be shown in
Section~\ref{perf-memory}.  To determine
whether it is safe to reclaim a certain node, the thread scans the
hazard pointers of all other threads to check if one of them is
currently using it.  Since each thread has $k$ hazard pointers and can
hold $R$ removed elements in its private list, a crashed thread can
prevent only $k + R$ removed elements from being reclaimed. The HP
scheme thus bounds the amount of memory which can be occupied by
removed nodes, even in the presence of thread failures.

HP can be extended to support an arbitrary number of hazard pointers
per thread as explained by Michael~\cite{Michael:HazardPointers}, but
unfortunately, this change destroys the two important properties that
set HP apart from other reclamation schemes: Constant processing
time per element as well as the upper bound on unreclaimable nodes.
Aghazadeh et al.~\cite{Aghazadeh:2014:MOW:2611462.2611483} gave an
improved version of HP that reduces the number of comparisons per scan
to one, at the cost of increasing the amount of time between node
removal and node reclamation.

A proposal to add hazard pointers to the C++ standard library has been
brought up by Michael and Wong~\cite{Michael:2016:C++HP}, and is
currently receiving attention.

\subsection{Quiescent state based reclamation}

Quiescent State based Reclamation (QSR) is typically used to implement
read-copy-update (RCU) schemes~\cite{Mckenney98read-copyupdate,
  User-Level-RCU}.
It relies on the concept of a \emph{grace period} which is a time
interval $[a, b]$ such that, after time $b$, all nodes removed before
time $a$ can safely be reclaimed. QSR uses \emph{quiescent states} to
detect grace periods. A \emph{quiescent state} for some thread $T$ is
a state in which $T$ holds no references to shared nodes. In
particular, $T$ holds no references to any shared nodes which have
been removed from a lock-free data structure.  A time interval in
which every thread of the system has passed through at least one
quiescent state is therefore a grace period.

A typical way to implement QSR is by using a non-blocking \emph{fuzzy
barrier}~\cite{Gupta:1989:FBM:70082.68187}. 
The fuzzy-barrier is used to protect the code that performs the
reclamation. The threads try to enter the barrier and reclaim retired
nodes when they pass through a quiescent state.  In order to determine
whether all threads have reached the barrier (\ie whether they went
through at least one quiescent state) all threads have to be checked.
This incurs a performance overhead linear in the number of threads.

\subsection{Epoch based reclamation}

Epoch-based Reclamation (ER), introduced by Fraser~\cite{Fraser:Phd},
also relies on grace periods. Nodes that have been removed
from data structures are kept in thread-local \emph{limbo lists} that
hold the references to the nodes until it is safe to reclaim them. The
scheme uses three \emph{epochs} and each of the epochs has an associated
limbo list.

In ER, the programmer has to identify \emph{critical regions}
in which threads are allowed to access shared objects. These regions
have to be entered and left explicitly. A global \emph{epoch count} is
used to determine when no stale references exist to any object in a
limbo list.

Every thread has a flag that indicates whether this thread is
currently in a critical region as well as a local epoch count that
identifies the epoch in which it currently executes (in case it is
inside a critical region). The thread's local epoch count may lag at
most one epoch behind the global epoch.  Each time a thread enters a
critical region, it sets the flag and \emph{observes} the current
epoch, \ie it updates its local epoch to match the global epoch.  A
thread that removes a node from a data structure places this node on
the current limbo list that is associated with the
current epoch.

After some predetermined number of critical region entries, a thread
will attempt to update the global epoch. This succeeds only if all
threads in a critical region have already observed the current epoch
which can again be detected with a fuzzy barrier. In that case, the
limbo list that was populated \emph{two} epochs ago can safely be
reclaimed and the list itself recycled and reused for the next
epoch. Thus only three epochs (and limbo lists) are required in total.

To determine whether all threads have observed the new global
epoch, all thread-local epochs have to be checked. This incurs a
performance overhead linear in the number of threads. 

\subsection{New epoch based reclamation}

New Epoch-based Reclamation (NER) is an extension to ER proposed by
Hart et al.~\cite{Hart:2007:PMR:1316099.1316427}. The original
description of ER defines a critical region around every
operation. However, entering a critical region requires a sequentially
consistent memory fence and such operations can be expensive. This is
necessary to guarantee that another thread that tries to update the
global epoch actually sees the new value and therefore recognizes that
this thread is inside a critical region. Without this guarantee, a
race condition can occur, where the global epoch gets updated which in
turn allows a node to be freed even though it is still in use by some
thread, just because the update of this thread's critical region flag
was not noticed by the thread that updated the global epoch. In ER
every single operation on some lock-free data structure is
encapsulated in its own critical region, thus every such operation
requires a memory fence.

Hart et al.~\cite{Hart:2007:PMR:1316099.1316427} showed this overhead
for every single operation to be very significant. As a remedy, 
NER allows critical regions to cover several operations. For example,
when a group of operations on some data structure has to be performed
together, the critical region is entered before the first operation
and left after the last one, effectively expanding the region over all
operations and thus distributing the overhead for the region entry
over the whole group of operations. The drawback is that due to the
larger regions the global epoch might be updated less frequently which
could delay reclamation and thus increase memory pressure.

\subsection{Stamp-it}
\label{Stamp-it}

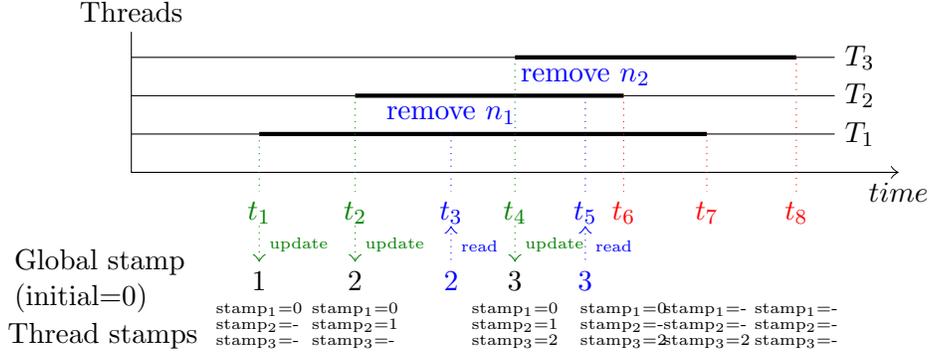
\begin{figure*}
\centering
\begin{tikzpicture}[xscale=8.5,yscale=1.7]
\draw[->] (-0.1,1.1) node[above]{Threads} -- (-0.1,0) -- (1.1,0) node[below] {$time$};

\draw               (-0.1,  0.3) -- (1.0, 0.3) node[right] {$T_1$};
\draw [ultra thick] (0.1,  0.3) -- (0.8, 0.3);

\draw               (-0.1, 0.6) -- (1.0, 0.6) node[right] {$T_2$};
\draw [ultra thick] (0.25, 0.6) -- (0.67, 0.6);

\draw               (-0.1, 0.9) -- (1.0, 0.9) node[right] {$T_3$};
\draw [ultra thick] (0.5, 0.9) -- (0.94, 0.9);

\node [align=left, left] at (0, -0.85) {Global stamp\\(initial=0)};
\node [align=left, below left] at (0.02, -1.1) {Thread stamps};

\draw [dotted, mygreen] (0.1,  0.3) -- (0.1, -0.15) node[below] {$t_1$};
\draw [dotted, mygreen, ->] (0.1, -0.42) -- (0.1, -0.7) node[above right,font=\fontsize{6}{6}\selectfont] {update};
\node                at (0.1, -0.85) {1};
\node [align=left,font=\fontsize{6}{6}\selectfont] at (0.1,  -1.2) {stamp$_1$=0\\stamp$_2$=-\\stamp$_3$=-};

\draw [dotted, mygreen]     (0.25,  0.6)  -- (0.25, -0.15) node[below] {$t_2$};
\draw [dotted, mygreen, ->] (0.25, -0.42) -- (0.25, -0.7) node[above right,font=\fontsize{6}{6}\selectfont] {update};
\node                    at (0.25, -0.85) {2};
\node [align=left,font=\fontsize{6}{6}\selectfont] at (0.25,  -1.2) {stamp$_1$=0\\stamp$_2$=1\\stamp$_3$=-};

\node [above, blue]   at (0.4,  0.3) {remove $n_1$};
\draw [dotted, blue]     (0.4,  0.3)  -- (0.4, -0.15) node[below] {$t_3$};
\draw [dotted, blue, <-] (0.4, -0.42) -- (0.4, -0.7) node[above right,font=\fontsize{6}{6}\selectfont] {read};
\node [blue]          at (0.4, -0.85) {2};

\draw [dotted, mygreen]     (0.5,  0.9)  -- (0.5, -0.15) node[below] {$t_4$};
\draw [dotted, mygreen, ->] (0.5, -0.42) -- (0.5, -0.7) node[above right,font=\fontsize{6}{6}\selectfont] {update};
\node                    at (0.5, -0.85) {3};
\node [align=left,font=\fontsize{6}{6}\selectfont] at (0.5,  -1.2) {stamp$_1$=0\\stamp$_2$=1\\stamp$_3$=2};

\node [above, blue]   at (0.61,  0.6) {remove $n_2$};
\draw [dotted, blue]     (0.61,  0.6)  -- (0.61, -0.15) node[below] {$t_5$};
\draw [dotted, blue, <-] (0.61, -0.42) -- (0.61, -0.7) node[above right,font=\fontsize{6}{6}\selectfont] {read};
\node [blue]          at (0.61, -0.85) {3};

\draw [dotted, red]     (0.67,  0.6) -- (0.67, -0.15) node[below] {$t_6$};
\node [align=left,font=\fontsize{6}{6}\selectfont] at (0.67,  -1.2) {stamp$_1$=0\\stamp$_2$=-\\stamp$_3$=2};

\draw [dotted, red]     (0.8,  0.3) -- (0.8, -0.15) node[below] {$t_7$};
\node [align=left,font=\fontsize{6}{6}\selectfont] at (0.8,  -1.2) {stamp$_1$=-\\stamp$_2$=-\\stamp$_3$=2};

\draw [dotted, red]     (0.94, 0.9) -- (0.94, -0.15) node[below] {$t_8$};
\node [align=left,font=\fontsize{6}{6}\selectfont] at (0.94,  -1.2) {stamp$_1$=-\\stamp$_2$=-\\stamp$_3$=-};
\end{tikzpicture}
\caption{Example showing global and local stamps of an execution with Stamp-it. Thick lines mark critical regions.}
\label{fig:stamp-it-concept}
\end{figure*}

We now introduce our new scheme, \emph{Stamp-it}. It is conceptually
similar to NER and therefore provides the many of same properties As
in ER/NER, the programmer has to define \emph{critical regions} that
are entered and left explicitly. A thread is only allowed to access
shared objects inside such regions.

When a thread enters a \emph{critical region} it increments a
\emph{global stamp} using an atomic fetch-and-add (FAA) and stores the
returned stamp in a thread-local data structure visible to the other
threads.  By setting the stamp in the data structure, the thread also
signals to other threads that it is now inside a \emph{critical
  region}.  When a thread retires a node for reclamation it takes the
current value of the \emph{global stamp}, stores it in a special field
of the node, and appends the node to the end of a thread-local retire
list. The node can be reclaimed as soon as all the threads that were
inside a critical region at the time the node was added to the retire
list have left their respective critical region.

When a thread leaves a critical region, it resets its stamp and tries
to reclaim retired nodes from the local retire list.  For that, it
must determine the \emph{lowest} stamp value of threads that are inside a
critical region, \ie the stamp value of the thread that has entered
a critical region at the earliest. Any node in the retire list that
has a stamp value that is less or equal to this lowest stamp can
safely be reclaimed. Since retired nodes are appended to the end of
the retire list they are strictly ordered by their stamp value.
Reclamation starts with the node with the lowest stamp and can stop as
soon as a node with a stamp higher than the current lowest stamp is
found.  No time is wasted on nodes that cannot yet be
reclaimed. Figure~\ref{fig:stamp-it-concept} illustrates this.

The initial value of the global stamp is zero. When thread $T_1$
enters its critical region at time $t_1$ it increments the global
stamp and stores the old value in its local stamp. The same happens
when $T_2$ enters its critical region at $t_2$ and $T_3$ at $t_4$. At
$t_3$, thread $T_1$ removes the node $n_1$ from some data structure
and marks it for reclamation. To that end, it reads the current value
of the global stamp, which is two since time instant $t_2$, stores
this value in the node and adds it to the local retire-list.
The node can be reclaimed once all threads that were
in a critical region at the time the node was marked ($t_3$) have left
their respective critical region. This can be determined by checking
if any thread in a critical region has a local stamp value that is
less than the node's stamp.
For the node $n_1$ this would be $t_7$ and for node $n_2$ it
would be $t_8$.

A straightforward implementation of this scheme is quite simple, but
will have runtime complexity linear in the number of threads since
all threads have to be scanned in order to determine the lowest
stamp. To improve this, we use a data structure that 
supports the following operations efficiently:
\begin{enumerate}
\item Add an element and assign a stamp to it (\texttt{push}). Stamps have to be
  strictly increasing, but not necessarily consecutive.
\item Remove a specific element, return \stamptrue if this element was
  the one with the lowest stamp (\texttt{remove}).
\item Get the highest stamp ever assigned to an element.
\item Get the lowest stamp of all elements.
\end{enumerate}
In addition, a global retire-list is introduced. It is used to collect
nodes that could not be reclaimed when their owning thread left its
critical region. The responsibility to reclaim these nodes is deferred
to the ``last'' thread as explained below.

Stamp-it uses this data structure as follows. Upon entering a critical
region the thread adds itself to the data structure, and gets a new
stamp value, defining a total order on the entries to the critical
regions.

When a thread retires a node, it requests the highest stamp from the
data structure, stores it in the node and appends the node to the end
of its local retire-list. If this pushes the number of entries in the
local retire-list over a certain threshold it immediately performs a
reclaim operation.  The reclaim operation requests the lowest stamp
from the data structure and reclaims all entries from the local
retire-list with a stamp value less than the requested one.

Upon leaving a critical region the thread removes itself from the data
structure and performs a reclaim operation on the local
retire-list. If the remove operation returns \stampfalse and the
number of nodes in the local retire-list exceeds some threshold, the
thread pushes all remaining entries to the global retire-list as an
ordered sublist. If the remove operation returns \stamptrue, \ie the
thread had the smallest stamp and was therefore ``lagging behind'' the
most and blocking reclamation, it will perform a reclaim operation on
the global retire-list. In contrast to the local retire-list, the
global retire-list is not totally ordered and therefore does not seem
to provide the same runtime guarantees. However, since it is organized
as a list of sorted sublists, each sublist needs to be scanned only up
to the node which has a stamp that is larger than or equal to the
lowest stamp returned. Therefore, if we maintain additional links from
sublist to sublist, the resulting total runtime is $O(n+m)$ where $n$ is
the total number of reclaimable nodes and $m$ is the number of ordered
sublists in the global retire-list.

We implemented the data structure as a lock-free doubly-linked list
based on the proposal by Sundell and Tsigas~\cite{SundellTsigas:2008}.
This data structure maintains sentinel \emph{head} and \emph{tail}
nodes which are used to store the highest and lowest stamp values,
respectively.  The \texttt{push} operation first increments the head's
stamp using an atomic fetch-and-add (FAA), stores the returned value
in the node it is currently inserting and then tries to insert the
node into the linked list, right after the head, using an atomic
compare-and-swap (CAS) operation.  The \texttt{remove} operation
unlinks the node from both directions, and returns \stamptrue if the
node was last, \ie the tail's predecessor.

Every thread holds a thread-local control block that is used as a node
in this list. A thread that enters a critical region simply calls
\texttt{push} with its node. Thus, the linked list in direction from
tail (smallest stamp) to head (largest stamp) defines the order in
which the threads have entered their respective critical regions.
When a thread leaves its critical region it calls remove. If the
return value is \stamptrue, it first updates tail's stamp to match the
value of the new predecessor, and then it performs a reclaim operation
on its local retire-list as well as the global retire-list. Otherwise,
the thread performs a reclaim operation on its local retire-list, and
if the number of remaining nodes exceeds some threshold, it moves the
remaining local list to the global retire-list.

The algorithm is clearly lock-free. In the absence of contention,
entering a critical section takes constant time, and leaving a
critical section time proportional to the number of reclaimable
nodes. The time per node is therefore amortized constant. In
Section~\ref{Stamp-it-perf}, we experimentally show that even under
load, the number of retry iterations is small (constant).

\subsection{DEBRA}
DEBRA (Distributed Epoch Based Reclamation), introduced by
Brown~\cite{Brown:2015:RML:2767386.2767436}, is an adaptation of ER. The
main difference is that its operations perform in only $O(1)$ steps. This
is achieved by incrementally scanning the flags of all the other threads
when entering a critical region. With each critical region entry only one
thread is checked, thus the cost of scanning $n$ threads is amortized over
$n$ enter operations. However, it still has to scan all threads.

We included DEBRA in this section for completeness, but we have not yet
implemented it, and therefore we did not consider it in the experimental
analysis. However, this is planned for future work; the results in this
paper will be updated accordingly.

\section{Experimental setup}
\label{sec:experiments}

We evaluate the described memory reclamation schemes with respect to
various factors. The tests are set up similarly to those performed by
Hart et al.~\cite{Hart:2007:PMR:1316099.1316427} and we also repeated
most of those analyses. This section shows only a subset of the results,
the remaining results can be found in the Appendix. All results including
the raw data and the scripts that were used are available on GitHub
(\url{https://github.com/mpoeter/emr-benchmarks}). This section provides
details on all aspects of our experiments.

\begin{table}
	\caption{The four machines used in the experimental evaluation}
\vspace{3mm} 
	\label{tbl:Machines}
\begin{tiny}
	\begin{tabular}{l|l|l|l|l}
		& \textbf{AMD} & \textbf{Intel} & \textbf{XeonPhi} & \textbf{Sparc} \\ \hline
		CPUs &
			\parbox[t]{3cm}{4x AMD Opteron(tm) \\ Processor 6168} &
			\parbox[t]{3cm}{8x Intel(R) Xeon(R) \\ CPU E7 - 8850 @ 2.00GHz} &
			\parbox[t]{3cm}{1x Intel(R) Xeon Phi(TM) \\ coprocessor x100 family} &
			4x SPARC-T5 \\ \hline
		Cores/CPU & 12 & 10 & 61 & 16 \\ \hline
		SMT & - & 2 & 4 & 8 \\ \hline
		Hardware Threads & 48 & 160 & 244 & 512 \\ \hline
		Memory & 128 GB & 1 TB & 16 GB & 1 TB \\ \hline
		OS &
			\parbox[t]{3cm}{Linux 4.7.0-1-amd64 \#1 SMP \\ Debian 4.7.6-1 (2016-10-07) \\ x86\_64 GNU/Linux} &
			\parbox[t]{3cm}{Linux 4.7.0-1-amd64 \#1 SMP \\ Debian 4.7.6-1 (2016-10-07) \\ x86\_64 GNU/Linux} &
			\parbox[t]{3cm}{Linux 2.6.38.8+mpss3.8.1 \#1 SMP \\ Thu Jan 12 16:10:30 EST 2017 \\ k1om GNU/Linux} &
			\parbox[t]{3cm}{SunOS 5.11 11.3 \\ sun4v sparc sun4v} \\ \hline
		Compiler &
			\parbox[t]{3cm}{gcc version 6.3.0 20170205 \\ (Debian 6.3.0-6)} &
			\parbox[t]{3cm}{icpc version 17.0.1 (gcc \\ version 6.0.0 compatibility)} &
			\parbox[t]{3cm}{icpc version 17.0.1 (gcc \\ version 5.1.1 compatibility)} &
			\parbox[t]{3cm}{gcc version 6.3.0 (GCC)}
	\end{tabular}
\end{tiny}
\end{table}

\subsection{Implementation}

The tests, data structures and reclamation schemes have been
implemented in C++11/14, using an adapted version of the interface
proposed by Robison~\cite{Robison:2013}. This proposal introduces the
concept of a \texttt{guard\_ptr} which allows a thread to get a safe
reference to a shared node, \ie the \texttt{guard\_ptr} ensures that
the node cannot be reclaimed as long as the \texttt{guard\_ptr}
instance exists.  Extending this interface, we introduce the concept
of a \texttt{region\_guard}. This is used in the implementations of
NER, QSR and Stamp-it to associate critical regions with the scope of
\texttt{region\_guard} instances. This reduces the costs of
\texttt{guard\_ptr} instances created inside the scope of a
\texttt{region\_guard}.  In NER and Stamp-it, a critical region thus
spans the lifetime of any \texttt{guard\_ptr} or
\texttt{region\_guard} instance. Since QSR is considered to be inside
a critical region at all times, each thread can go through a quiescent
state once the last \texttt{guard\_ptr} or \texttt{region\_guard}
instance is released.

\subsection{Benchmarks}

We tested the reclamation schemes on a (1) queue, a (2) linked-list
and a (3) hash-map.  The queue is based on Michael and Scott's
design~\cite{Michael:1996:SFP:248052.248106}, the linked-list and
hash-map are based on Michael's improved
version~\cite{Michael:2002:HPD:564870.564881} of Harris' list-based
set~\cite{Harris:2001:PIN:645958.676105}.  The List and Queue
benchmarks have a parameter to control the \emph{number of elements
  initially} in the data structures. For the List benchmark the key
range is calculated to be twice the initial list size. The
probabilities of inserting and removing nodes are equal, keeping the
size of the list and queue data structures roughly unchanged
throughout a given run. The List benchmark has a \emph{workload
  parameter} that determines the fraction of updates (remove/insert)
of the total number of operations. A workload of $0\%$ corresponds to
a search-only use case, while a workload of $100\%$ corresponds to an
update-only use case.

Our experiments are \emph{throughput oriented} in the following sense.
The main thread spawns $p$ child threads and starts a timer. Every
child thread performs operations on the data structure under scrutiny
until the timer expires. Upon timer expiry the child threads are
stopped and the parent thread calculates the average execution time
per operation by summing up the runtime of each child thread and its
number of performed operations.

Each benchmark was performed with 30 trials, with eight seconds
runtime per trial. Most of the benchmarks focus on \emph{performance},
and calculate the \emph{average runtime per single operation} for each
trial.  Each thread calculates its average operation runtime by
dividing its active, overall runtime by the total number of operations
it performed. The total average runtime per operation is then
calculated as the average of these per-thread runtime values. 

It is important to note that all 30 trials were performed sequentially
within the same process. This is especially important in case of the
HashMap benchmark as the hash-map is retained over the whole
runtime. This means that a result calculated in the first trial can be
found in the hash-map and reused in a subsequent trial. For this
reason, performance will be worse at the beginning, while the hash-map
is in the ``warm up phase'', but will improve over time when it
becomes filled and more items can be reused. But also in the other
benchmarks, it is possible that previous trials have impact on
later ones, \eg due to an already initialized memory manager. It was
a deliberate design decision to run all trials in the same process as
this might more closely reflect a real world situation.

The Queue and List benchmarks are synthetic micro-benchmarks, exactly
as used by Hart et al.~\cite{Hart:2007:PMR:1316099.1316427}. The
HashMap benchmark is intended to highlight other properties of the
reclamation schemes. It mimics the calculation in a complex simulation
where partial results are stored in a hash-map for later reuse. These
partial results are relatively large, so in order to limit the total
memory usage the number of entries in the hash-map is kept below some
threshold by evicting old entries using a simple FIFO policy. The
resulting benchmark has the following properties:
\begin{itemize}
\item there is no upper bound on the number of nodes that are
  \emph{intentionally} blocked from reclamation.
\item the average lifetime of each \texttt{guard\_ptr} is relatively
  long.
\item the memory footprint of each node is significant, putting
  additional pressure on the reclamation scheme to reclaim nodes
  efficiently and in a timely manner.
\end{itemize}
Since there is no upper bound on the number of nodes that need to be
available for a thread, the standard HP scheme is insufficient; thus
an extended version has to be used that allows a dynamic number of
hazard pointers as explained by Michael~\cite{Michael:HazardPointers}.
The number of buckets in the hash-map is 2048, the maximum number of
entries in the hash-map is 10000. There are 30000 possible partial
results and every thread has to calculate or reuse 1000 partial
results per ``simulation''.  The size of a partial result is 1024
bytes.

Last but not least, the GuardPtr benchmark is used to
measure the base cost/overhead of creating and destroying
\texttt{guard\_ptr} instances. Each thread repeatedly creates a
\texttt{guard\_ptr} on a shared node and immediately destroys it; no
other operations are performed.

\subsection{Environment}

We executed our tests on four machines with different
(mi\-cro)ar\-chitectures.  Their respective specifications are shown
in Table~\ref{tbl:Machines}. These machines all have a relatively
large number of cores and hardware supported threads, allowing us to
run our experiments at a scale not found in most prior studies.  We
did not experiment with oversubscribed cores.

On Sparc we used \texttt{jemalloc}~\cite{Evans06ascalable} since in
Solaris the \texttt{libc} implementation of \texttt{malloc} and
\texttt{free} uses a global lock. We did not use \texttt{libumem} (a
scalable memory manager that is part of all newer Solaris
versions\footnote{\url{https://blogs.oracle.com/ahl/number-11-of-20:-libumem}}),
because we ran into some sporadic but severe performance drops when
running with a very large number of threads ($>200$). We suspect these
issues to be caused by large numbers of cross-thread deallocations. As
alternatives we tried Hoard~\cite{Berger:2000:HSM:384264.379232} and
\texttt{jemalloc}, but Hoard showed similar symptoms as
\texttt{libumem} while \texttt{jemalloc} did not.

ER/NER try to advance the epoch every 100 critical section entry.
In the List and Queue benchmarks, a \texttt{region\_guard} spans 100
benchmark operations, so this is the size of the critical region for
QSR, NER and Stamp-it. QSR executes the fuzzy barrier when it exits
the critical region. In HPR the local retire list is scanned once its
threshold is exceeded; the threshold is $100 + \sum_{i=0}^p K_i*2$
where $p$ is the number of threads and $K_i$ is the number of hazard
pointers for the thread with index $i$

\section{Experimental Results}

In this section we present the results of a subset of our experiments.
First, we show that Stamp-it meets the expectations
with respect to the expected average runtime complexity. We then
present thread scalability results, and finally investigate
the reclamation efficiency for all described schemes.
All results including the data are available on GitHub
(\url{https://github.com/mpoeter/emr-benchmarks}).

\subsection{Stamp-it base performance}
\label{Stamp-it-perf}

\begin{figure*}
  \centering
  \includegraphics[width=\textwidth]{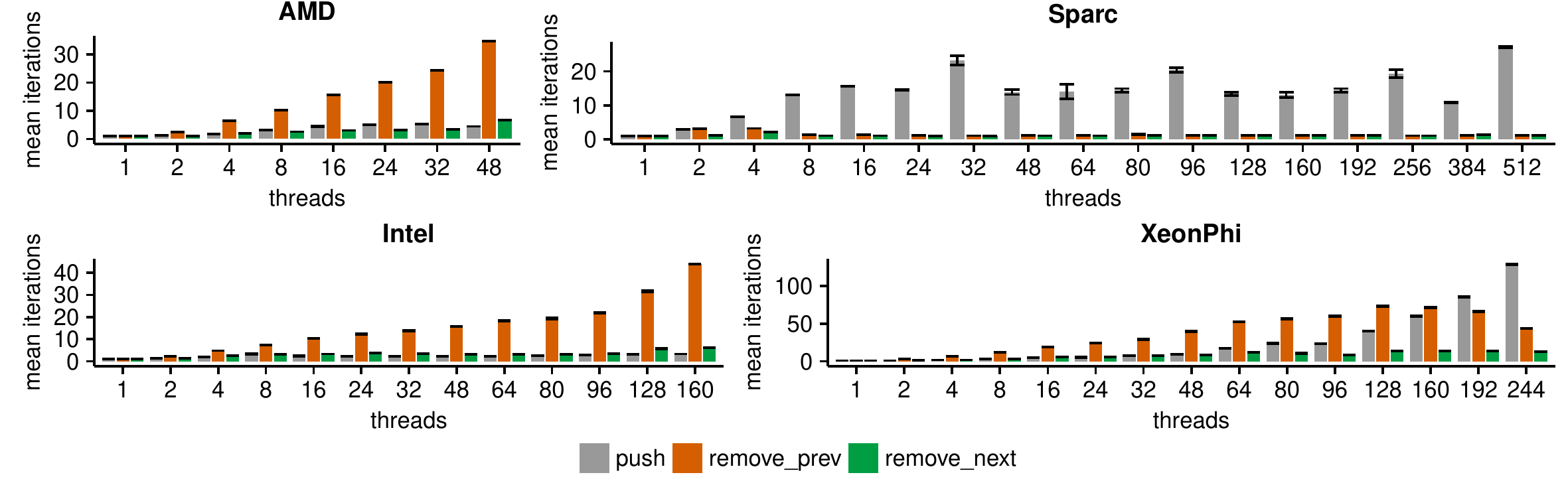}
  \caption{Mean number of iterations for the respective operations in the GuardPtr benchmark.}
  \label{fig:benchmark-Stamp-it-guard_ptr}
\end{figure*}
\begin{figure*}
  \centering
  \includegraphics[width=\textwidth]{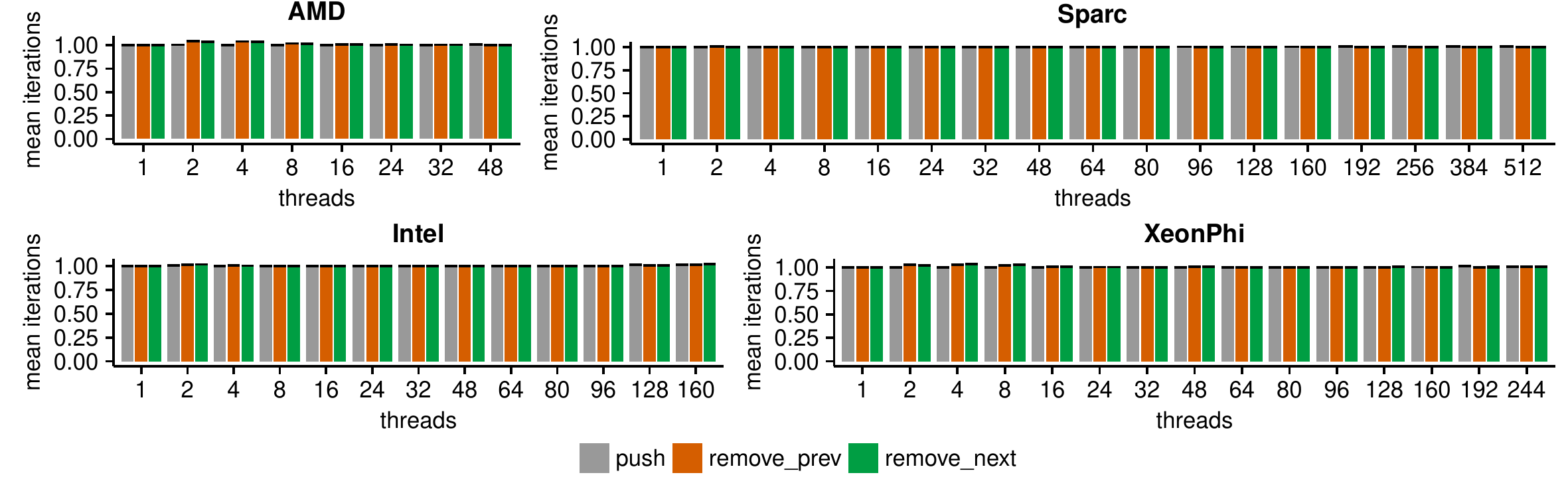}
  \caption{Mean number of iterations for the respective operations in the Queue benchmark.}
  \label{fig:benchmark-Stamp-it-queue}
\end{figure*}
\begin{figure*}
  \centering
  \includegraphics[width=\textwidth]{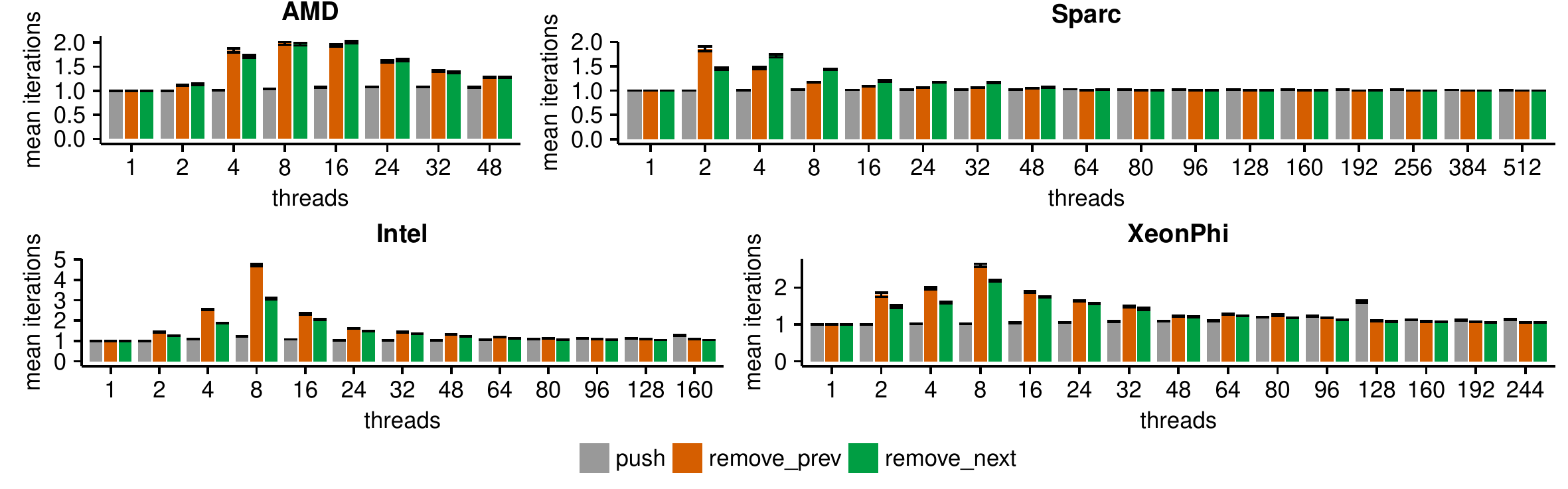}
  \caption{Mean number of iterations for the respective operations in the HashMap benchmark.}
  \label{fig:benchmark-Stamp-it-hash_map}
\end{figure*}

We first give an experimental analysis of the effective average number
of steps for the data structure operations in Stamp-it. To that end,
we use thread-local performance counters to keep track of the number
of retries due to failed CAS operations in \texttt{push} and
\texttt{remove}, thus allowing to calculate the average number of
iterations per operation. Since the data structure is based on a
doubly linked list, the \texttt{remove} operation builds on two other
operations \texttt{remove\_from\_prev} and \texttt{remove\_from\_next}
to remove the node from both directions~\cite{SundellTsigas:2008}.

The benchmarks were run as described in Section~\ref{sec:experiments},
but instead of average time per operation, the average number of
iterations in \texttt{push}, \texttt{remove\_from\_prev} and
\texttt{remove\_from\_next} has been measured. The results for the
various benchmarks are shown in
Figures~\ref{fig:benchmark-Stamp-it-guard_ptr},
\ref{fig:benchmark-Stamp-it-queue} and
\ref{fig:benchmark-Stamp-it-hash_map}. The results for the List
benchmark are omitted because they a qualitatively similar to those
from the Queue benchmark. However, they can be found in 
Appendix~\ref{appendix:stamp-it-list-benchmark}.

The GuardPtr benchmark is the most interesting, since this is kind of
a ``stress test'', \ie it simulates the worst case scenario where all
threads just insert and immediately remove themselves from the data
structure. Essentially, this scenario tests the scalability of the
data structure itself. As can be seen in
Figure~\ref{fig:benchmark-Stamp-it-guard_ptr}, the average number of
iterations is less than the number of threads in all cases, suggesting
that even in this worst case scenario the expected average runtime
complexity is $O(p)$.  

Interestingly, the behavior differs significantly on the
various architectures. For AMD, Intel and the XeonPhi the results are
dominated by the number of iterations in
\texttt{remove\_from\_prev}. On XeonPhi the number of iterations in
\texttt{push} increases significantly once the number of threads is
greater than 120. The reason for this could lie in the SMT based
architecture with 61 physical cores and the way instructions are
scheduled \cite{XeonPhi-Microarchitecture}. For SPARC the situation is
completely opposite: The number of threads has almost no impact on
the number of iterations in the remove-methods, instead the number of
iterations in \texttt{push} is increasing, but quite unsteadily.

It is likewise interesting to see how the data structure performs
under ``normal'' conditions. As can be seen in
Figures~\ref{fig:benchmark-Stamp-it-queue}
and~\ref{fig:benchmark-Stamp-it-hash_map}, which give results from the
Queue and HashMap benchmarks, the number of threads has almost no
measurable impact on the number of iterations for all three
methods: Numbers are more or less constant, with a few
outliers in the HashMap benchmark were we can see a a small increase
around 4-16 threads that again decreases with a growing number of
threads.
 
\subsection{Base costs}

\begin{figure*}
  \centering
  \includegraphics[width=\textwidth]{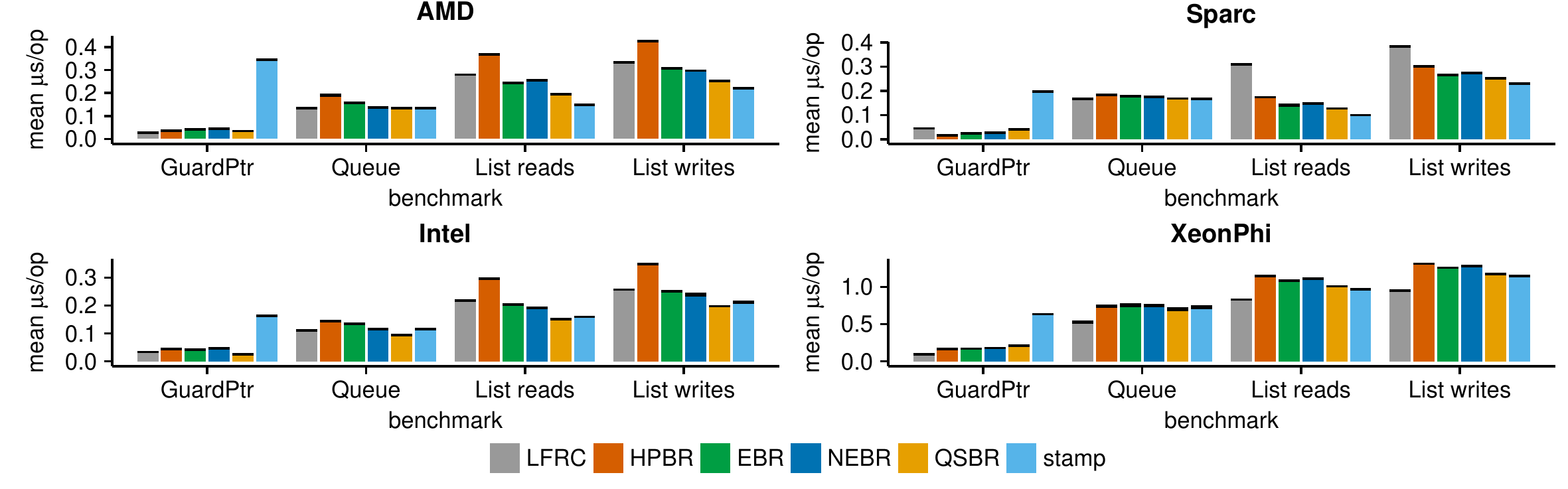}
  \caption{Base costs of the five plus one schemes in single thread runs.}
  \label{fig:base-costs}
\end{figure*}

In this analysis we measure the base costs of the schemes. We use a
single thread to eliminate contention on the used data structures, so
resulting performance differences are caused solely by creating and
destroying \texttt{guard\_ptr} instances. We also include the GuardPtr
benchmark to measure the pure overhead of creating and releasing
\texttt{guard\_ptr} instances without any other operations involved.

All benchmarks except HashMap were run on all machines using a single
thread, 30 trials and eight seconds runtime. The results are shown in
Figure~\ref{fig:base-costs}; ``List reads'' corresponds to the List
benchmark with a workload of 0\% (\ie read-only) and ``List writes''
corresponds to the List benchmark with a workload of 100\% (\ie all
operations are either insert or delete). The number of elements for
the List and Queue benchmarks was $10$.  The HashMap benchmark was
excluded here because it has a very high runtime dominated by the
simulated calculations; the overhead for allocating and releasing
\texttt{guard\_ptr}'s is rather irrelevant.

Stamp-it performs very poorly in the GuardPtr benchmark due to the
more expensive operations to insert and remove the thread from the
internal data structure. But the results show that there is hardly any
trace of this overhead in the other benchmarks; in some cases Stamp-it
is the fastest of all schemes.  This is due to the fact that, just
like NER and QSR, Stamp-it also uses the \texttt{region\_guard}
concept to amortize the cost of these insert and remove calls over a
larger number of operations.
We again observe significant
differences between Sparc and the Intel based architectures. On Sparc
LFRC is significantly slower then HP.

\subsection{Scalability with threads}
\label{thread-scalability}

\begin{figure*}[!tb]
  \centering
  \includegraphics[width=\textwidth]{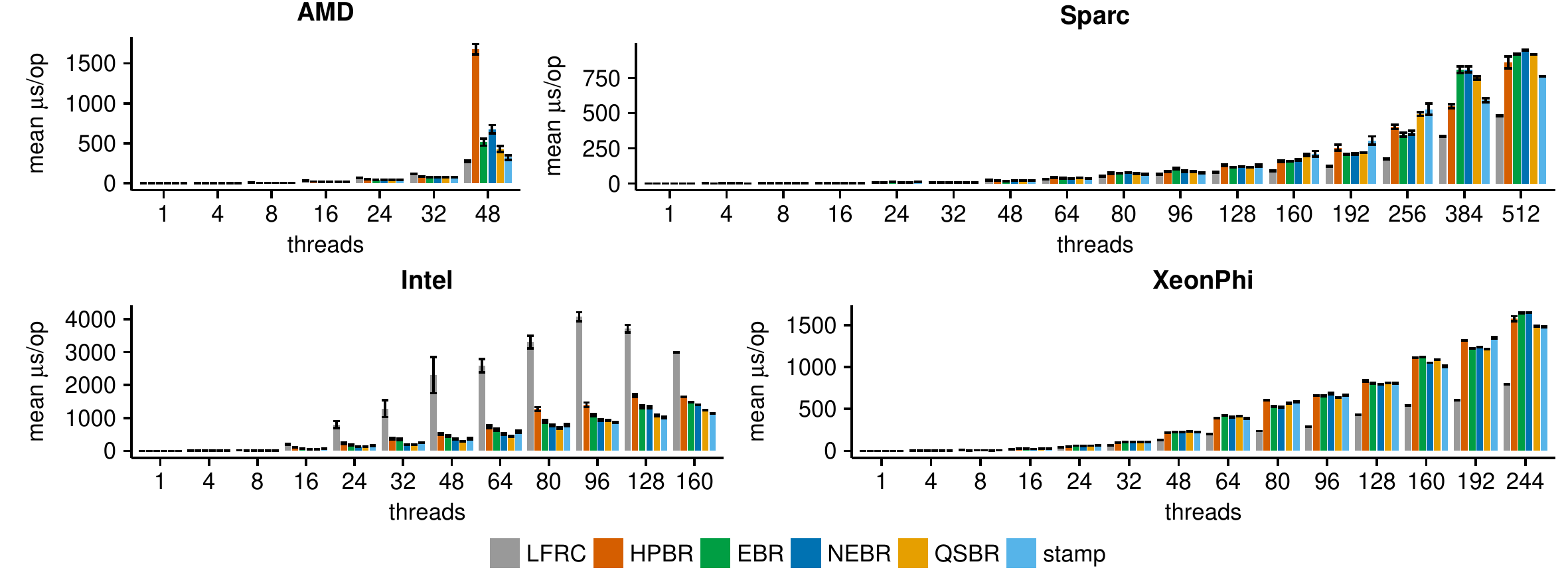}
  \caption{Performance of the Queue benchmark with varying number of threads.}
  \label{fig:threads-queue}
\end{figure*}
\begin{figure*}[!tb]
  \centering
  \includegraphics[width=\textwidth]{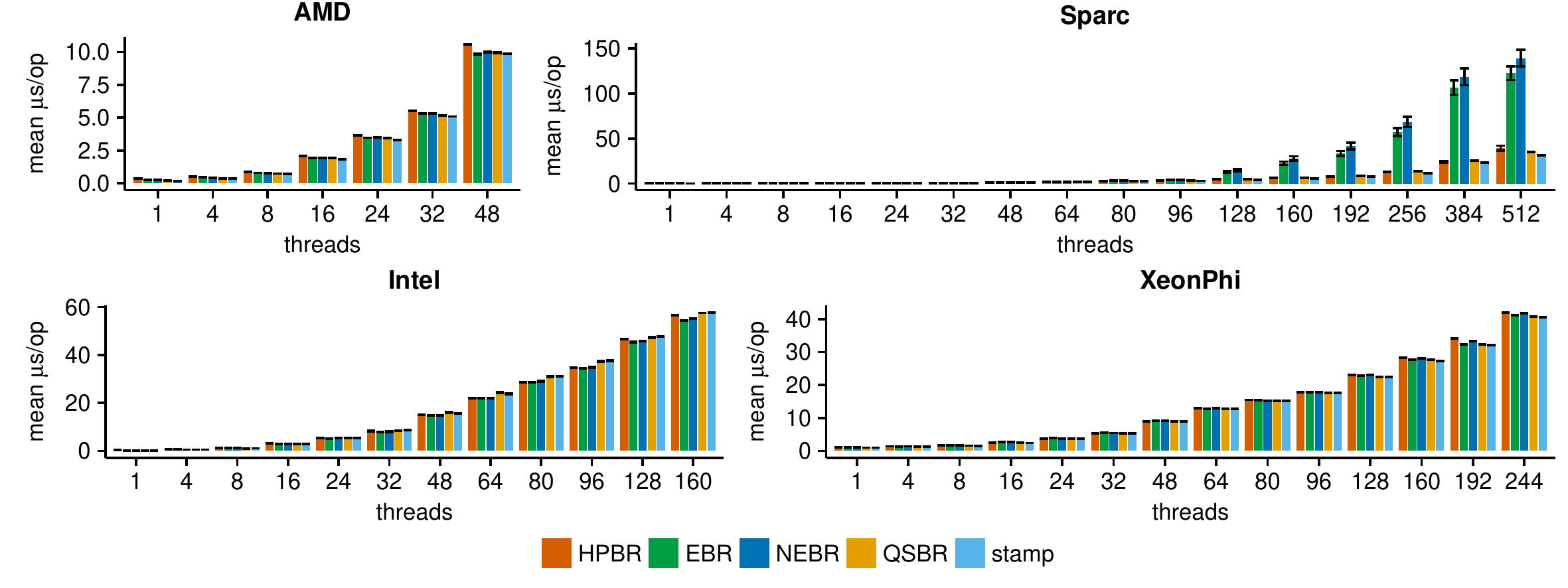}
  \caption{Performance of the List benchmark with 10 elements, a workload of 20\% and varying number of threads (without LFRC).}
  \label{fig:threads-list-20}
\end{figure*}
\begin{figure*}[!tb]
  \centering
  \includegraphics[width=\textwidth]{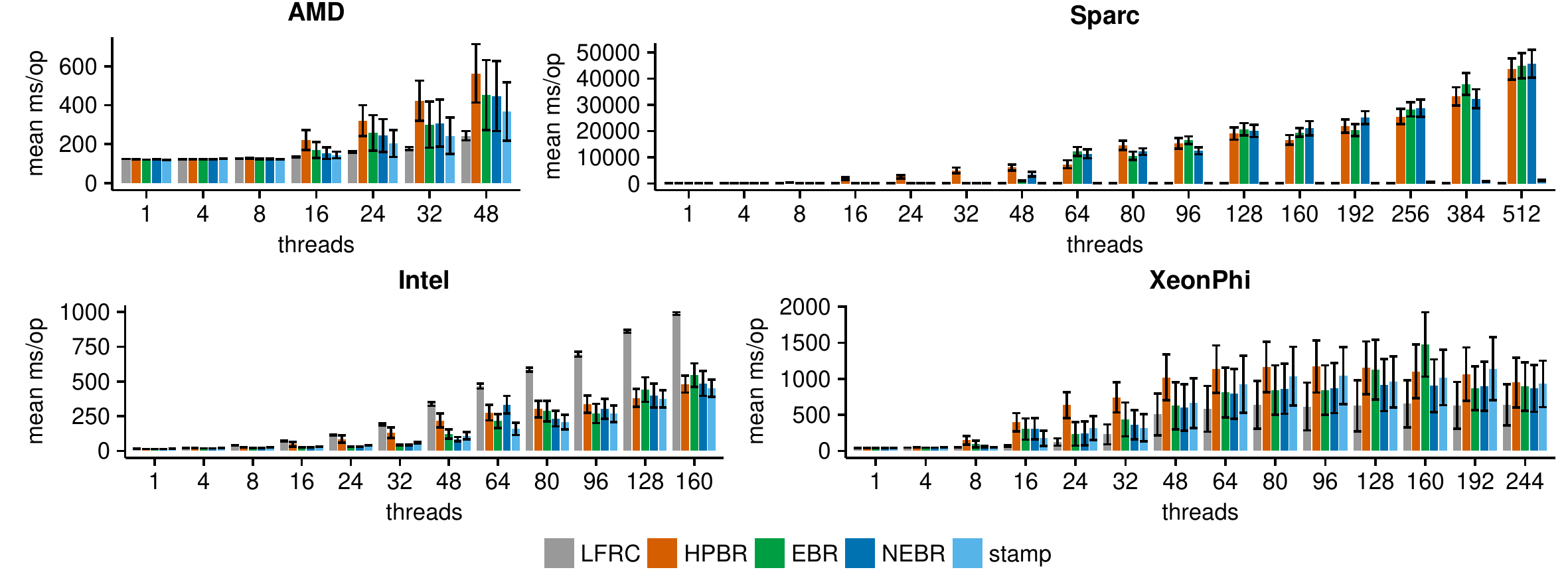}
  \caption{Performance of the HashMap benchmark with varying number of threads.}
  \label{fig:threads-hash_map}
\end{figure*}

We now study the effect of increasing the number of threads that
share a single instance of some data structure.

Figure~\ref{fig:threads-queue} shows the performance of the
reclamation schemes in the Queue benchmark. Surprisingly, LFRC
performs by far the best on Sparc and on XeonPhi, but is by far the worst on
Intel. On AMD, HP has a huge performance drop when running with the
maximum number of threads. A similar effect can be seen by the other
schemes as well, but much less significant.  Apart from these
exceptions, all schemes seem to scale largely equally well in this
scenario.

For the results of the List benchmark in
Figure~\ref{fig:threads-list-20}, LFRC has been excluded since it
performs exceedingly poor in this scenario, especially with a larger
number of threads. On AMD, Intel and XeonPhi, all schemes are more on
less on par, but on Sparc EB and NER show a significant degradation
when the number of threads grows beyond 128. What is surprising,
though, is that in all those cases NER performs consistently worse
than ER. This is quite unexpected, since NER was designed to have less
overhead than ER. We did not investigate the reasons for this in more
detail, but one assumption is that this might be caused by a larger
number of unsuccessful attempts to update the global epoch, which
could be caused by NER's dependence on larger critical regions.

Finally, the results for the HashMap benchmark are shown in
Figure~\ref{fig:threads-hash_map}.  QSR has been excluded because it
scales very poorly on all architectures in this update-heavy
scenario. On AMD, ER, NER and Stamp-it scale almost perfectly, while
LFRC's and HP's performance starts to degrade once the number of
threads grows beyond 16. On Intel, LFRC scales very poorly while all
other schemes scale more or less equally well, but not as well as
on AMD. On XeonPhi on the other hand, LFRC scales best while HP's
performance starts degrading with more than 16 threads, but it again
improves with more than 128 threads. The other schemes continuously
loose performance when the number of threads grows from 16 to
${\sim}80$, but then stays more or less the same.

The biggest surprise is the result on Sparc. Here, the performance of
HP, ER and NER degrades dramatically, while LFRC and Stamp-it scale
almost perfectly. With 512 threads the performance difference between
LFRC/Stamp-it and the other schemes is a factor of ${\sim}4000$. The
reason for this will become clear when we look at the results of the
reclamation efficiency analysis in the next section.

\subsection{Reclamation efficiency}
\label{perf-memory}

This analysis focuses on how efficiently (fast) the various schemes
actually reclaim retired nodes. An increased reclamation efficiency
can drastically reduce memory pressure, which in turn can have a
significant impact on the overall performance. Nonetheless, this
aspect is usually disregarded in analyses of concurrent reclamation
schemes.

\begin{figure*}[!tb]
  \centering
  \includegraphics[width=\textwidth]{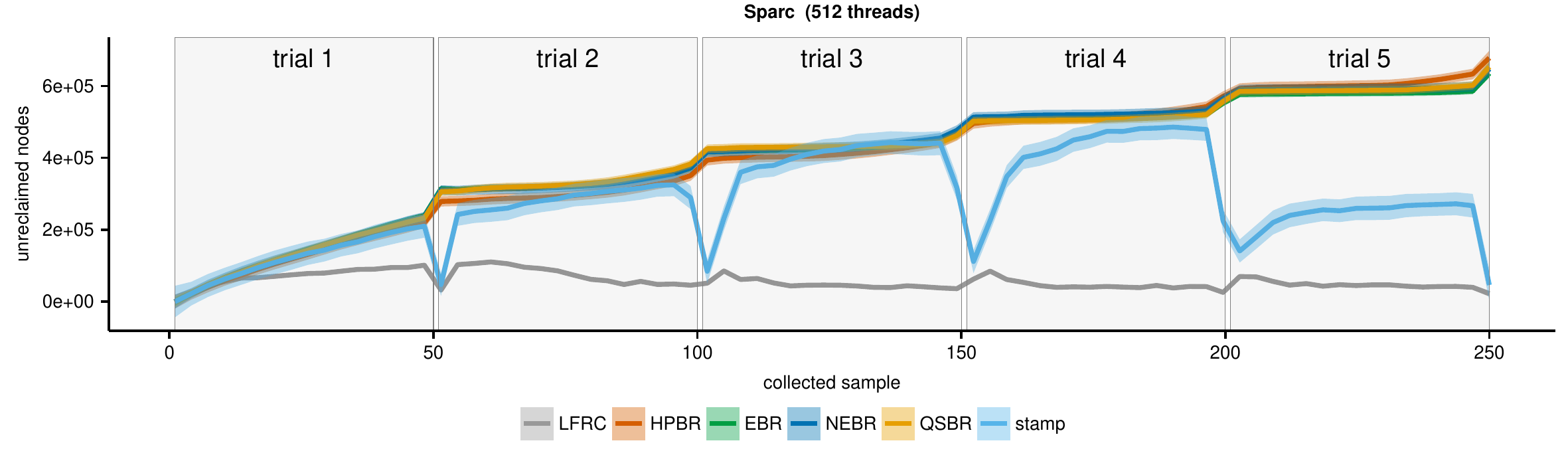}
  \caption{Number of unreclaimed nodes over time in the HashMap benchmark on Sparc. The $x$-axis is the current sample.}
  \label{fig:unreclaimed-nodes-hash_map-Sparc}
\end{figure*}

To measure reclamation efficiency we use thread-local performance
counters that track the number of allocated and reclaimed nodes. By
calculating the differences, we get the number of unreclaimed nodes,
which is our measurement for efficiency; a smaller number of
unreclaimed nodes means that the reclamation scheme works more
efficiently.

The plots in this analysis show the development of the number of
unreclaimed nodes over time. Each configuration is run with five
trials, each with a runtime of eight seconds. During each trial a
total of 50 samples are collected. Since the benchmarks are randomized
each configuration with the five trials is run 20 times to account for
any fluctuation in the measured samples. The plots show the smoothed
conditional means of the measured samples of those 20 runs over the
number of samples recorded during each run.

\begin{figure*}[!tb]
  \centering
  \includegraphics[width=\textwidth]{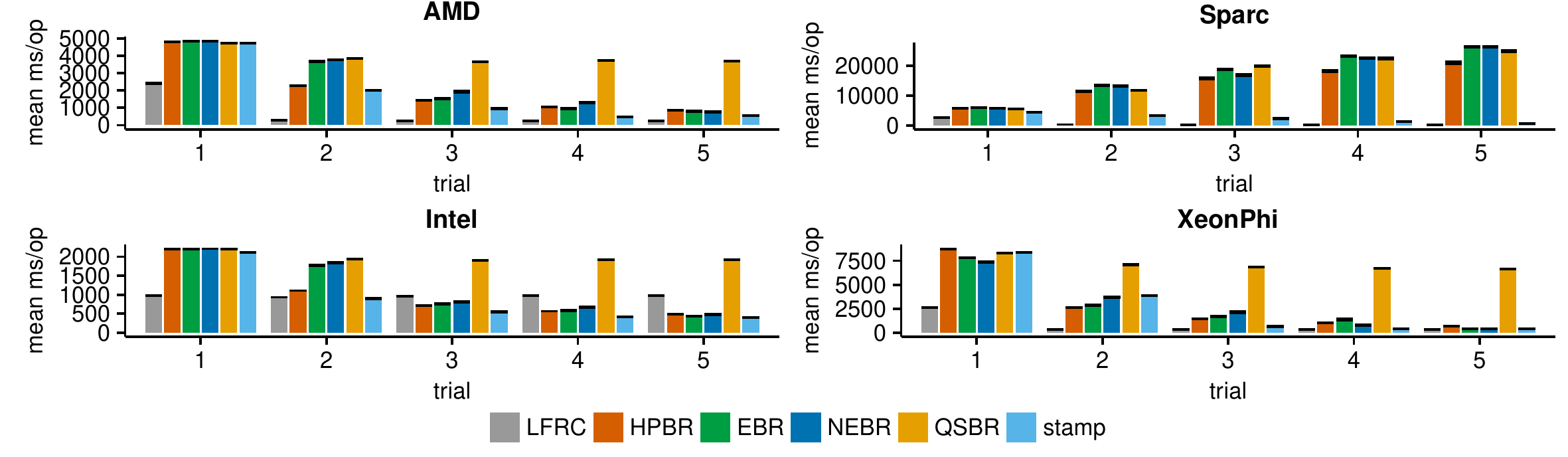}
  \caption{Development of runtime over time in the HashMap benchmark.}
  \label{fig:unreclaimed-nodes-hash_map-runtime}
\end{figure*}

For reclamation efficiency, reference counting is the ``gold
standard''. In contrast to all other schemes there is no delay: A node
is reclaimed immediately when the last thread drops its reference to
that node. So in all the plots, LFRC can bee seen as the baseline
against which all other schemes have to be measured. One has to keep
in mind, though, that LFRC is not a general reclamation scheme, since
the reclaimed nodes are not returned to the memory manager, but stored
in the internal free-list.

Figure~\ref{fig:unreclaimed-nodes-hash_map-Sparc} shows the results
for the HashMap benchmark on Sparc. The results for the other
benchmarks and machines can be found in
Appendix~\ref{appendix:reclamantion-efficiency}.

What can be seen is that the number of unreclaimed nodes for HP, ER,
NER and QSR is constantly increasing. It does not even go down at the
end of the trials when all threads are stopped.

When a thread terminates, all schemes add the remaining nodes to a
global list. But who is responsible to reclaim them, and when? In
Stamp-it the responsibility is transferred to the ``last''
thread. Other schemes do not have a notion of a ``last'' thread, so
the global retire-list is checked by each thread when it performs
reclamation on its local retire-list. When a thread tries to reclaim
nodes from the global list it steals the whole list, reclaims all
reclaimable nodes and then re-adds the remaining nodes to the global
list.  This leads to a race during the end of a trial; whoever steals
the list might not be able to reclaim all nodes yet, but when the
remaining nodes are re-added to the global list, there might be no
threads left. Stamp-it mitigates this race as it is cheap to check
whether the global stamp has changed since reclamation has started,
and so it can restart reclamation with the new stamp value. Obviously,
the effects of this race are more pronounced the more threads are
involved. The behavior in shown in
Figure~\ref{fig:unreclaimed-nodes-hash_map} in
Appendix~\ref{appendix:reclamantion-efficiency} is a direct result of
this race.

The failure to efficiently reclaim nodes increases memory pressure,
which has a direct impact on the
runtime. Figure~\ref{fig:unreclaimed-nodes-hash_map-runtime} shows the
development of the runtime over the five trials. On Sparc we can see
that the runtime of HP, ER, NER and QSR is increasing with each trial,
while LFRC and Stamp-it is decreasing. On the other architectures
runtime is decreasing for all schemes except QSR. This would be the
expected behavior since more results can be reused once the hash-map
has been filled.

HP also performed very poorly on the other architectures when the number of
threads becomes very large. This is caused by the larger threshold for number
of retired nodes to achieve amortized constant processing-time.

\section{Conclusion and Future Work}

This paper introduced \emph{Stamp-it}, a new, general purpose memory
reclamation scheme with attractive features. To the best of our knowledge,
this is the first non-reference counting based scheme that does not have to
scan all other threads to determine reclaimability of a node.

We have also presented a large scale experimental study, comparing the
performance of five plus one reclamation schemes on four different
architectures in various scenarios.  Our empirical results show that
Stamp-it matches or outperforms the other analyzed reclamation schemes
in almost all cases.

All of the analyzed schemes are implemented in portable, standard
conform C++, based on the standardized interface proposed by
Robison~\cite{Robison:2013}; the full source code is available on
GitHub (\url{https://github.com/mpoeter/emr}).

For future work we plan to add an implementation of
DEBRA~\cite{Brown:2015:RML:2767386.2767436} and include it in the
benchmark results. It might be interesting to look for other data
structures that could replace the doubly linked list, i.e., data
structures that have less overhead while providing all the required
properties. In this context we might also try to relax some of these
properties (e.g., use a partial order instead of a strict order for
thread entries) in order to reduce contention on the data structure.

\bibliographystyle{abbrv}
\bibliography{reclamation} 

\newpage
\appendix
\section{Additional results}

This appendix contains additional results, some of which were briefly
discussed but not shown in the main text.

\subsection{Lock-free Reference counting}
\label{LFRC-performance}

Reference counting is prone to false sharing
as the reference counter is part of the node,
which can usually be avoided by extra \emph{padding}, however, at the
cost of a higher memory overhead.

Another reason for the high overhead of LFRC is the global free-list
that is shared by all threads and can lead to high contention. A
simple way to reduce this contention is to use fixed-size thread-local
free-lists as buffers. Both improvements, padding as well as the
thread-local free-lists, have been implemented here.

LFRC is often criticized for its bad performance. As already mentioned, 
we tried to improve that by implementing the following two
extensions: a) padding to avoid false sharing between the
reference counter and other node members, and b) bounded, thread-local
free-lists. The following variations of LFRC were used in the experiment:
\begin{itemize}
\item unpadded -- LFRC without padding.
\item unpadded-20 -- like unpadded, but with local free-lists of 20 entries.
\item padded -- LFRC with padding
\item padded-20 -- like padded, but with local free-lists of 20 entries.
\end{itemize}

\begin{figure*}[!htb]
  \centering
  \includegraphics[width=\textwidth]{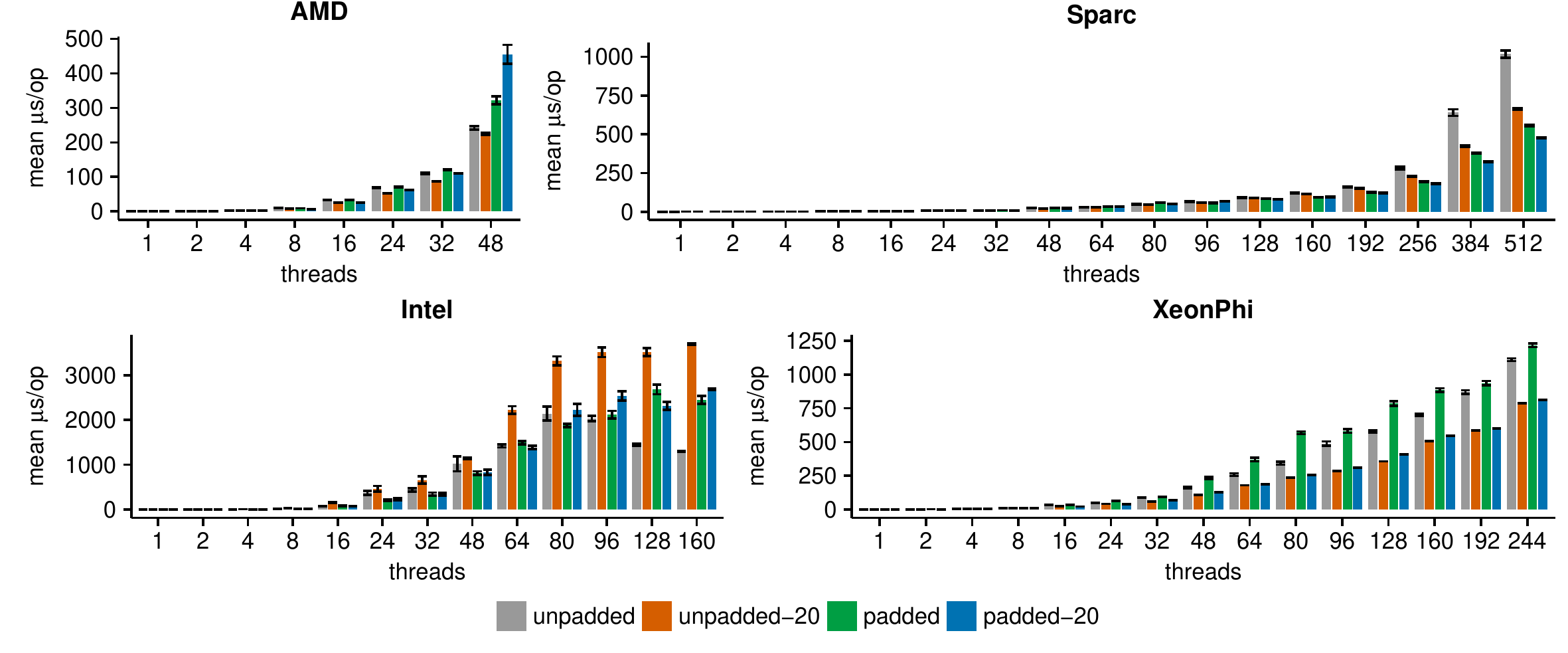}
  \caption{Performance of different LFRC configurations with varying number of threads in the Queue benchmark.}
  \label{fig:benchmark-LFRC-queue}
\end{figure*}

\begin{figure*}[!tb]
  \centering
  \includegraphics[width=\textwidth]{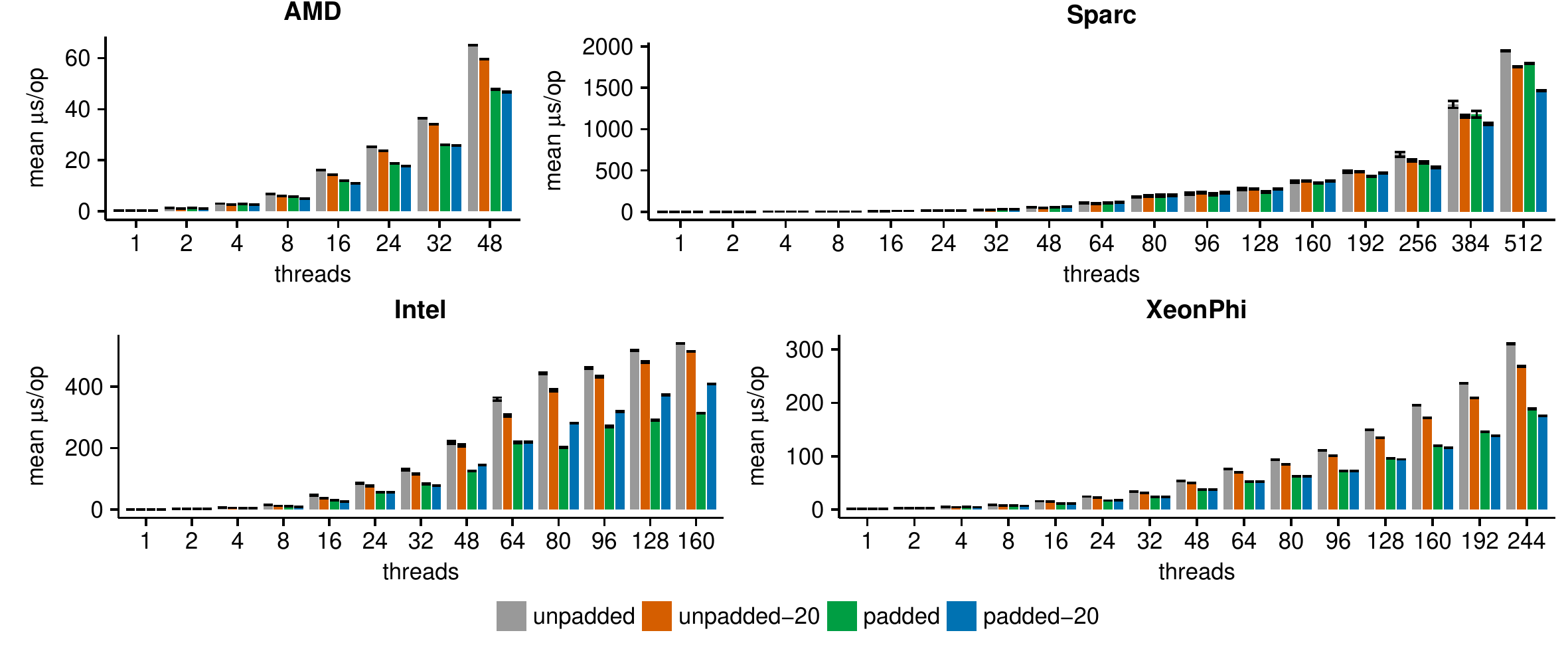}
  \caption{Performance of different LFRC configurations with varying number of threads in the List benchmark with a workload of 20\%.}
  \label{fig:benchmark-LFRC-list20}
\end{figure*}

\begin{figure*}[!tb]
  \centering
  \includegraphics[width=\textwidth]{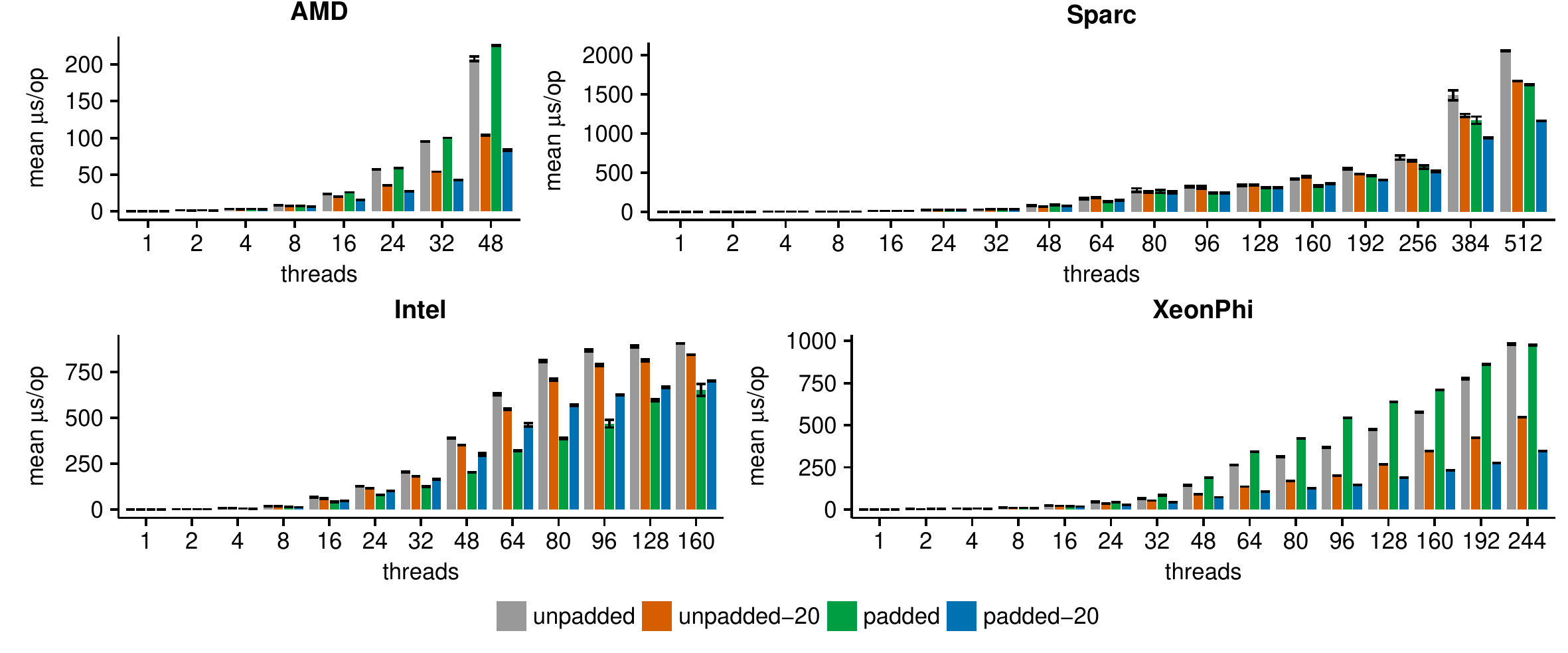}
  \caption{Performance of different LFRC configurations with varying number of threads in the List benchmark with a workload of 80\%.}
  \label{fig:benchmark-LFRC-list80}
\end{figure*}

\begin{figure*}[!tb]
  \centering
  \includegraphics[width=\textwidth]{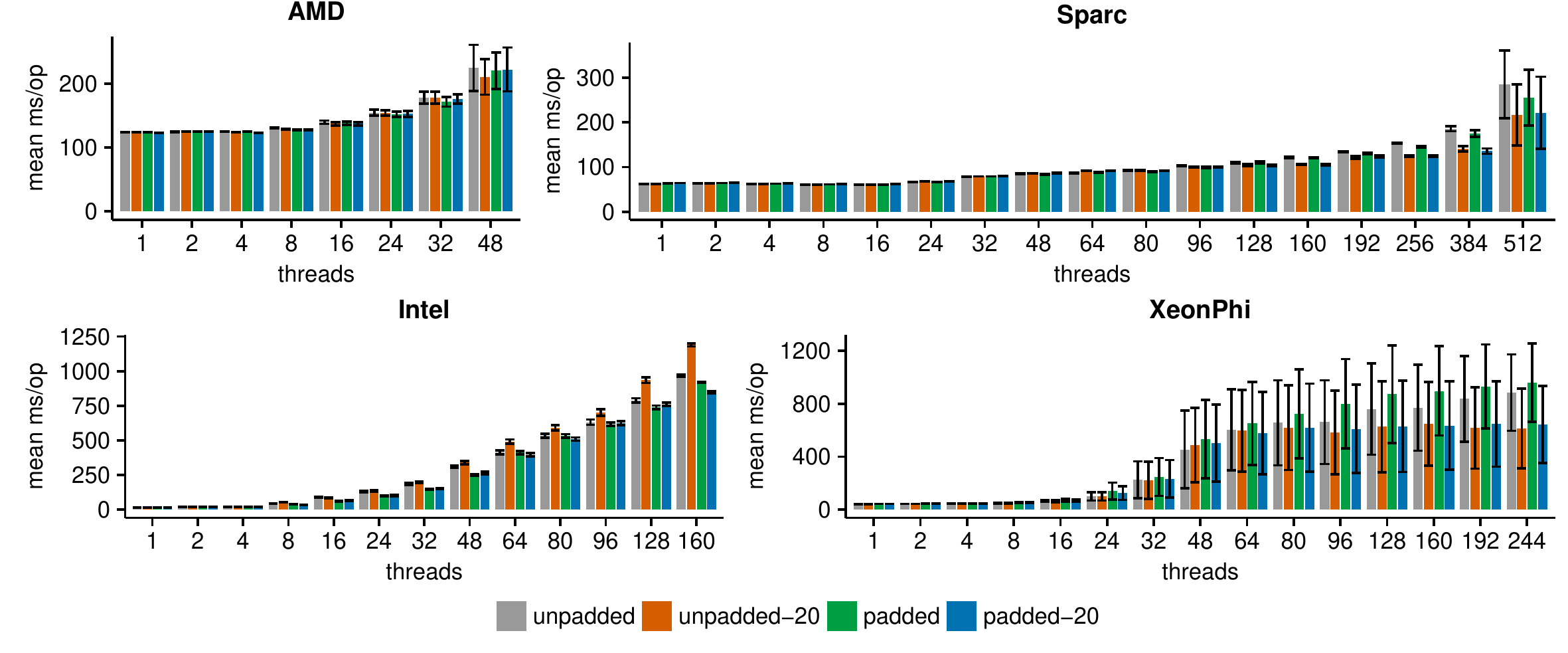}
  \caption{Performance of different LFRC configurations with varying number of threads in the HashMap benchmark.}
  \label{fig:benchmark-LFRC-hashmap}
\end{figure*}

The results are shown in the Figures~\ref{fig:benchmark-LFRC-queue},
\ref{fig:benchmark-LFRC-list20}, \ref{fig:benchmark-LFRC-list80}
and~\ref{fig:benchmark-LFRC-hashmap}.  The results are quite
interesting as there is no overall ``best'' configuration. Instead the
performance of the different configurations varies with both the
benchmark data structure and the CPU architecture. However, in almost
all cases at least one of the other configurations is significantly
faster than the original, unpadded LFRC.


\subsection{Stamp-it List benchmark}
\label{appendix:stamp-it-list-benchmark}

\begin{figure*}[!htb]
  \centering
  \includegraphics[width=\textwidth]{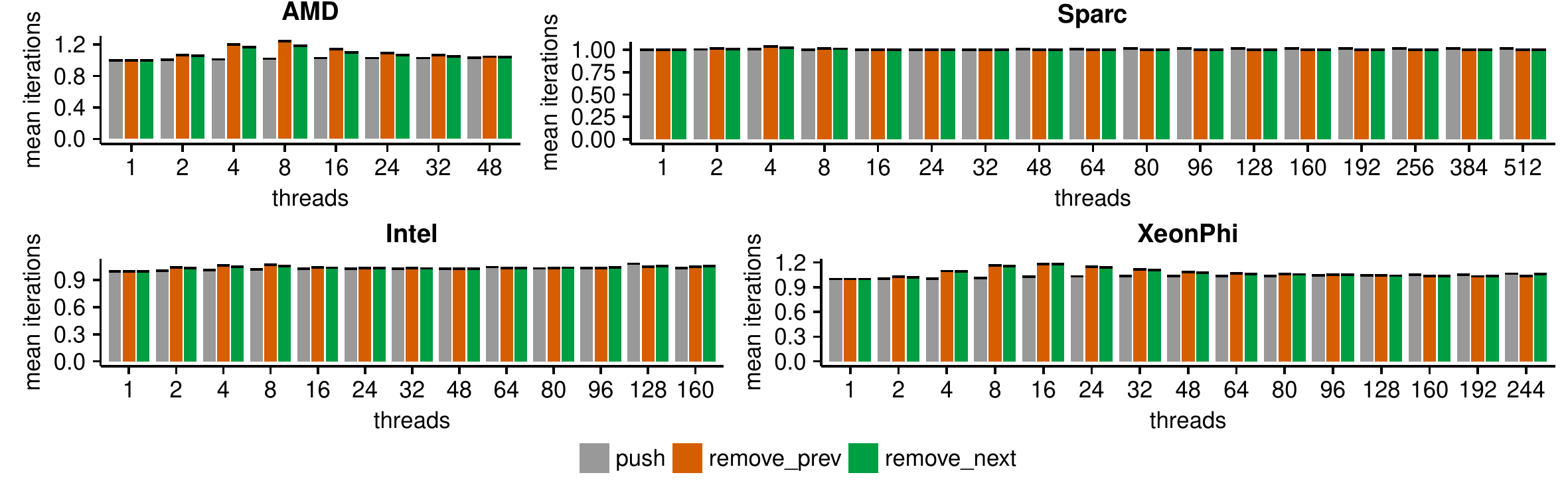}
  \caption{Mean number of iterations for the respective operations in the List benchmark with workload 20\%.}
  \label{fig:benchmark-Stamp-it-list20}
\end{figure*}

\begin{figure*}[!htb]
  \centering
  \includegraphics[width=\textwidth]{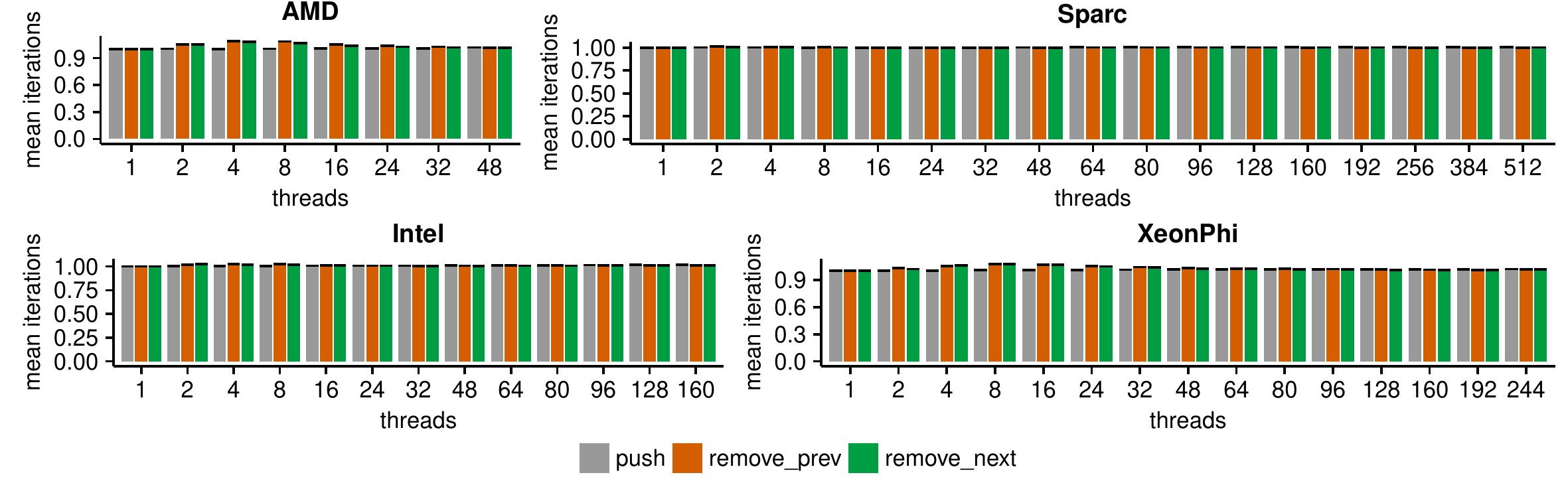}
  \caption{Mean number of iterations for the respective operations in the  List benchmark with workload 80\%.}
  \label{fig:benchmark-Stamp-it-list80}
\end{figure*}

Figures~\ref{fig:benchmark-Stamp-it-list20}
and~\ref{fig:benchmark-Stamp-it-list80} show the Stamp-it results from
the List benchmark with a workload of 20\% and 80\%. These results
were excluded from the main text since they are qualitatively similar
to the results from the Queue benchmark as shown in
Figure~\ref{fig:benchmark-Stamp-it-queue}.


\subsection{Scalability with workload}

This analysis which was omitted from the main text uses the List
benchmark to examine the workload's impact on the reclamation schemes,
by gradually increasing the read-to-update ratio of the performed
operations from read-only to update-only.

\begin{figure*}
  \centering
  \includegraphics[width=\textwidth]{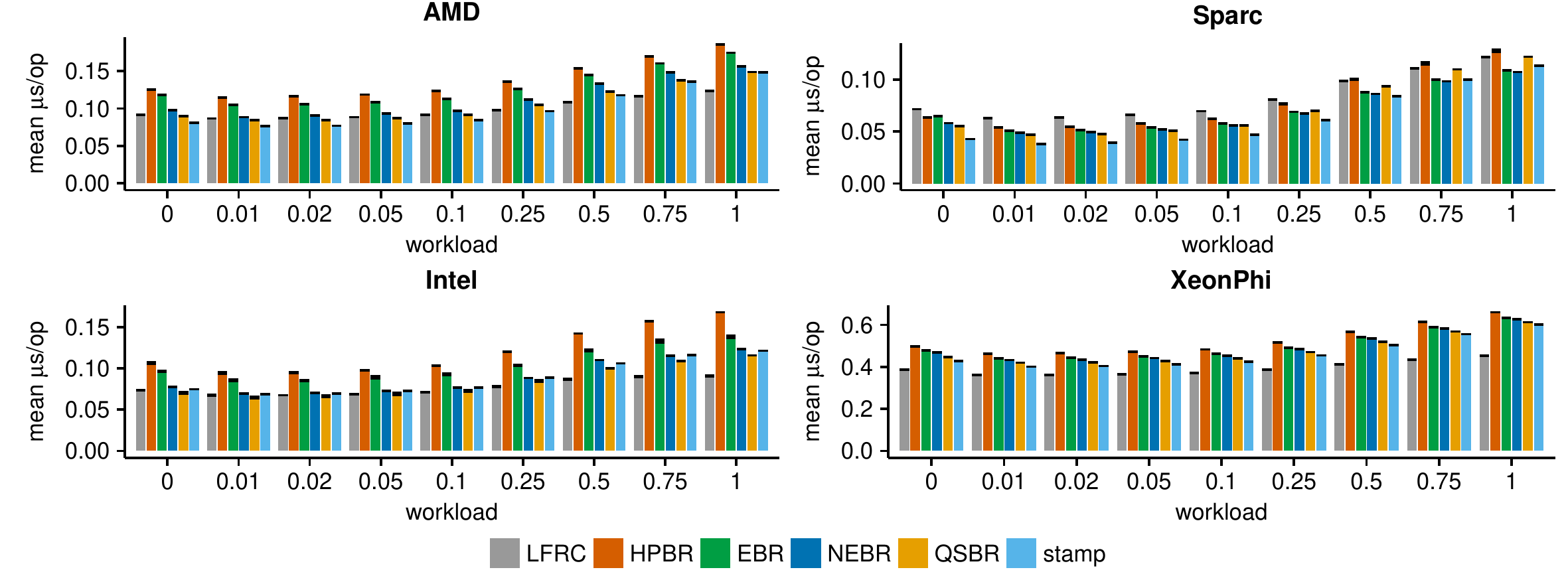}
  \caption{Effect of varying workload on a lock-free list with one element, one thread.}
  \label{fig:workload-1-elem-1-thread}
\end{figure*}

\begin{figure*}
  \centering
  \includegraphics[width=\textwidth]{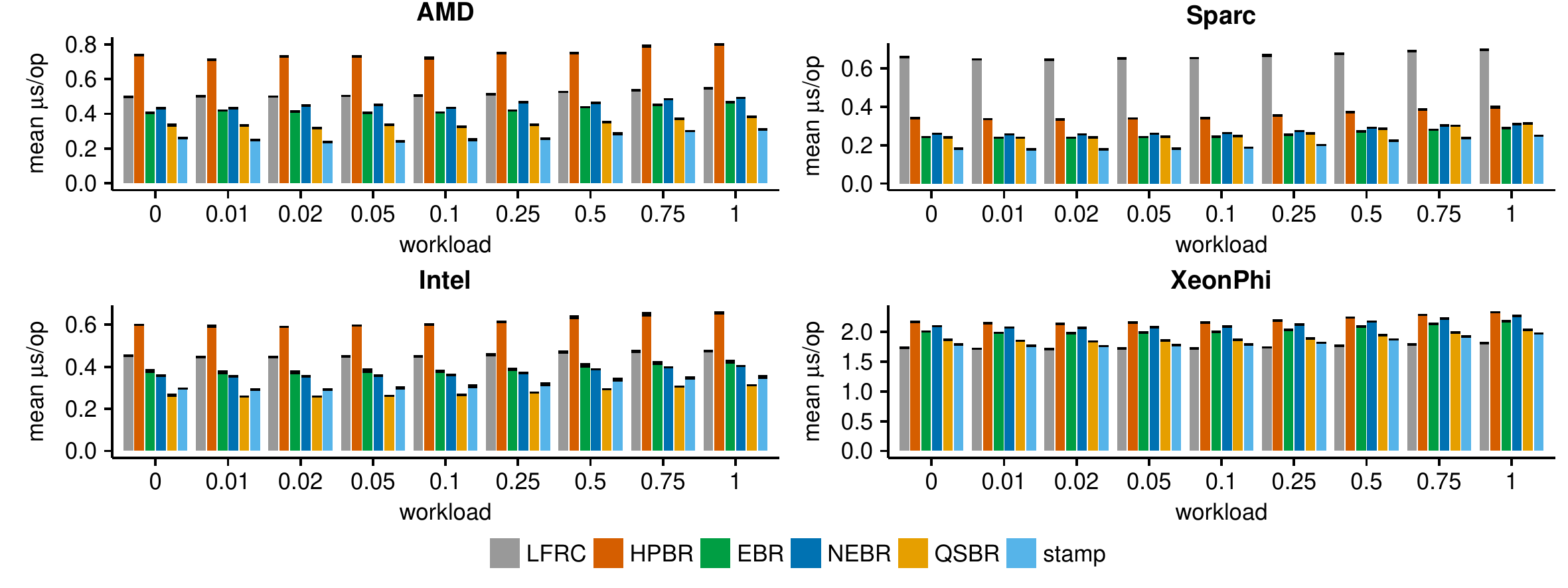}
  \caption{Effect of varying workload on a lock-free list with 25 elements, one threads.}
  \label{fig:workload-25-elem-1-thread}
\end{figure*}

\begin{figure*}
  \centering
  \includegraphics[width=\textwidth]{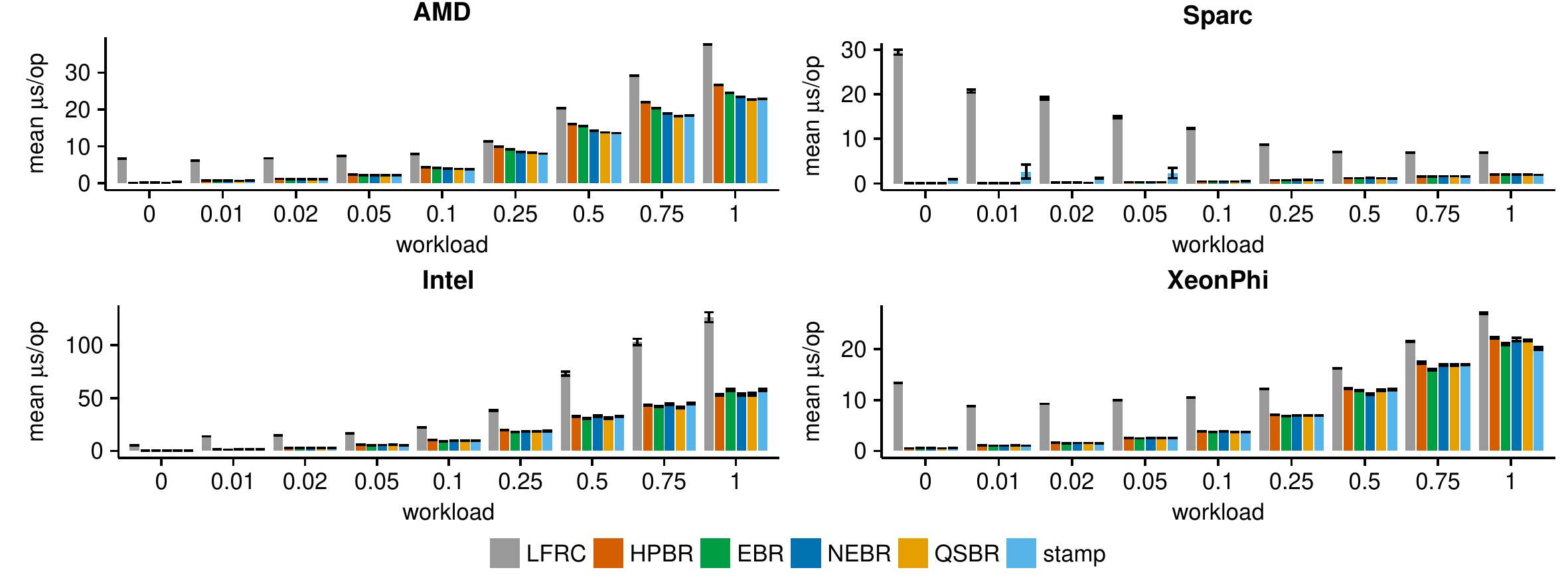}
  \caption{Effect of varying workload on a lock-free list with one element, 32 threads.}
  \label{fig:workload-1-elem-32-threads}
\end{figure*}

\begin{figure*}
  \centering
  \includegraphics[width=\textwidth]{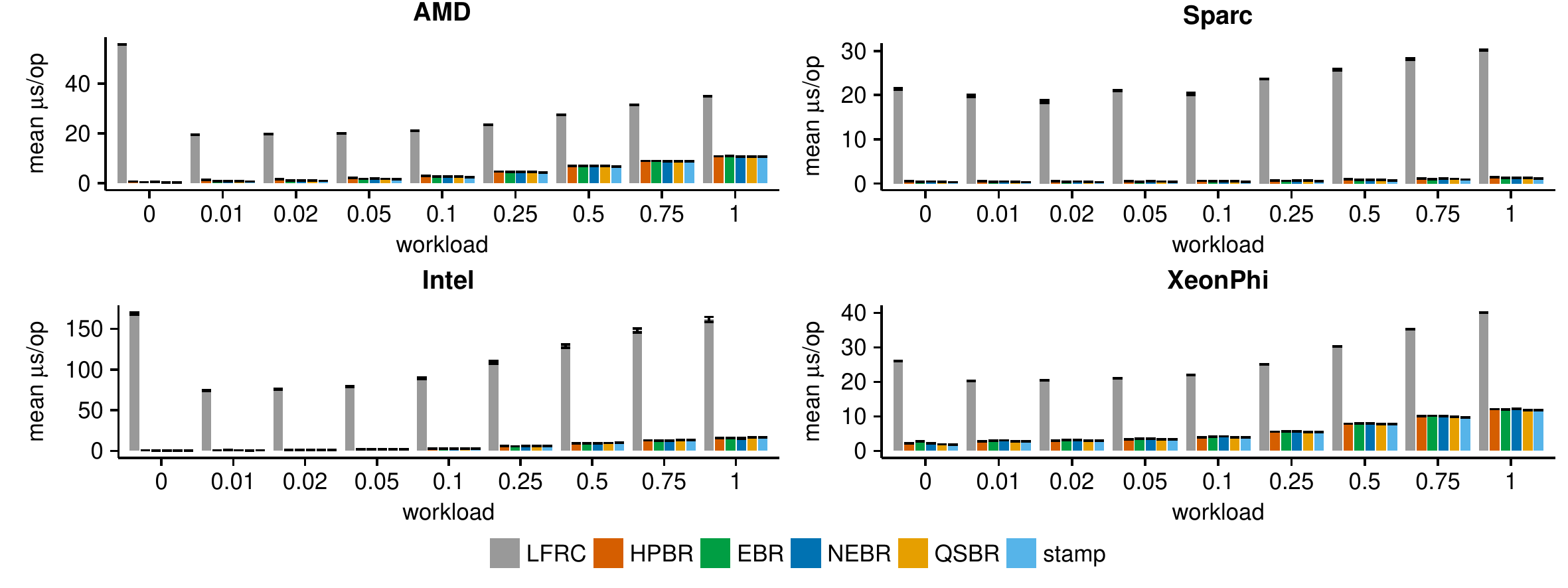}
  \caption{Effect of varying workload on a lock-free list with 25 elements, 32 threads.}
  \label{fig:workload-25-elem-32-threads}
\end{figure*}

\begin{figure*}
	\centering
	\includegraphics[width=\textwidth]{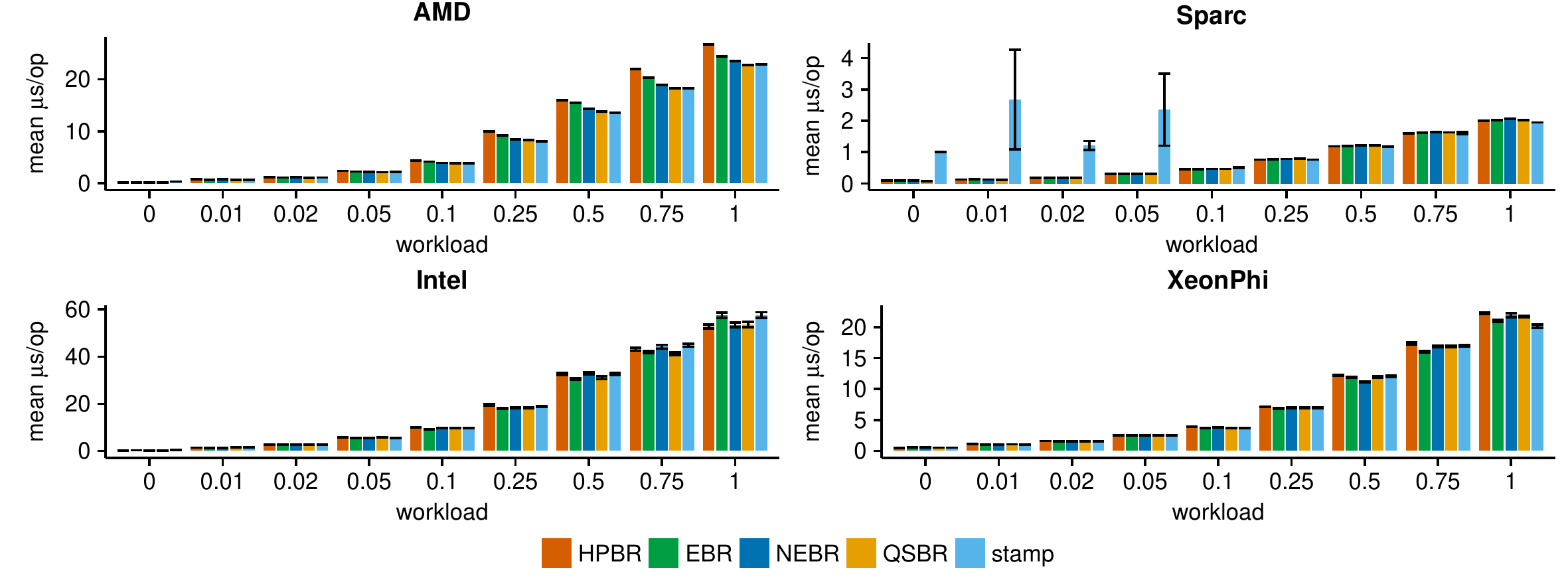}
	\caption{Effect of varying workload on a lock-free list with one element, 32 threads without LFRC.}
	\label{fig:workload-1-elem-32-threads-no-LFRC}
\end{figure*}

\begin{figure*}
	\centering
	\includegraphics[width=\textwidth]{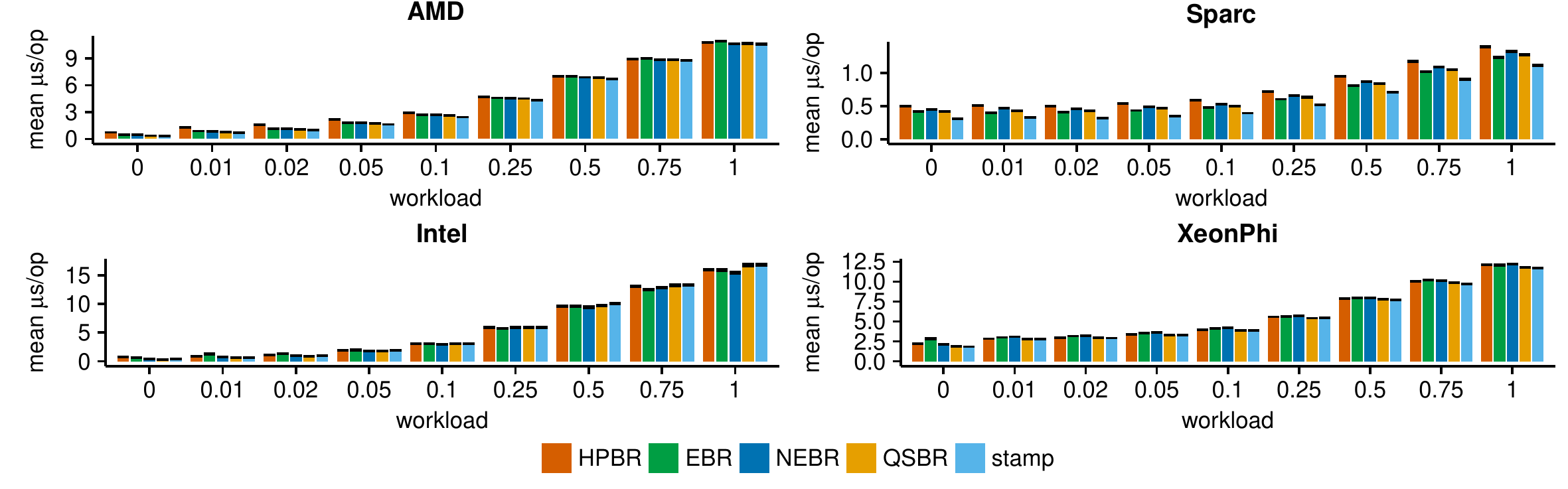}
	\caption{Effect of varying workload on a lock-free list with 25 elements, 32 threads without LFRC.}
	\label{fig:workload-25-elem-32-threads-no-LFRC}
\end{figure*}

When purely read-only operations are used no nodes get reclaimed, so
the schemes only differ in the performance overhead of acquiring and
releasing the necessary \texttt{guard\_ptr} instances. With an
increasing number of update operations, the performance overhead for
acquiring and releasing the \texttt{guard\_ptr} instances stays the
same (we still have to search the list the same way as for read-only
operations). But the more update operations are performed
(specifically delete operations), the more impact on the overall
performance is caused by the reclamation of retired nodes. The
benchmark was run in four different configurations:
\begin{itemize}
	\item one thread; one element (see Figure~\ref{fig:workload-1-elem-1-thread})
	\item one thread; 25 elements (see Figure~\ref{fig:workload-25-elem-1-thread})
	\item 32 threads; one element (see Figure~\ref{fig:workload-1-elem-32-threads} and~\ref{fig:workload-1-elem-32-threads-no-LFRC})
	\item 32 threads; 25 elements (see Figure~\ref{fig:workload-25-elem-32-threads} and~\ref{fig:workload-25-elem-32-threads-no-LFRC})
\end{itemize}
Each configuration was run with 30 trials and a runtime of eight seconds.
For LFRC the configuration with padding and a local free-list with 20
entries was used; based on the results from Section~\ref{LFRC-performance}
it seemed to be the overall best choice for this scenario.


As can be seen by the results of the various configurations, the workload
by itself seems to have no significant impact on the performance of the
reclamation schemes; within each configuration and architecture, all schemes
exhibit roughly the same slope, \ie the relative performance difference
between the schemes stays more or less the same, regardless of the workload.
Hart et al.~came to the same conclusion in their experiments~\cite{Hart:2007:PMR:1316099.1316427}.
This is not entirely unexpected, since insert and remove operations still
require the same lookup to be performed as in a search operation. The only
exception is LFRC, which actually shows a performance improvement on Sparc
in the configuration with one element and 32 threads (see
Figure~\ref{fig:workload-1-elem-32-threads}), but it starts out with a huge
gap to the other schemes. It is not entirely clear why LFRC can improve its
performance, but we suspect it is due to the way of how LFRC reuses reclaimed
nodes.

In the base cost analysis we saw that LFRC seems to incur a higher overhead on
the Sparc architecture. Figure~\ref{fig:workload-1-elem-1-thread} shows the
results for the configuration with one element and one thread. In this
configuration HP performs worst in almost all cases, while LFRC on the
other hand is almost always fastest, or at least on par with the fastest
scheme -- with the exception of Sparc, where LFRC performs worse than HP in
virtually all scenarios. This pattern also emerges from the results of all
other configurations, which corroborates the observation from the base cost
analysis that LFRC performs worse on Sparc, and is thus less well suited for
this architecture.

Naturally, LFRC performs significantly worse with a growing number of
threads as can be seen in Figures~\ref{fig:workload-25-elem-1-thread}
and~\ref{fig:workload-25-elem-32-threads}. What is quite interesting,
though, is that in the scenario with a single element (see
Figure~\ref{fig:workload-25-elem-1-thread}), on Sparc the performance
of LFRC is actually \emph{increasing} with a higher workload; the
other schemes and architectures do not show such an effect. Since
these results are dominated by the rather bad performance of LFRC,
Figures~\ref{fig:workload-1-elem-32-threads-no-LFRC}
and~\ref{fig:workload-25-elem-32-threads-no-LFRC} show the same
results with LFRC excluded.


What can be seen in Figure~\ref{fig:workload-1-elem-32-threads-no-LFRC} is
that in the configuration with 32 threads and a single element, in the first
scenarios, which have a low workload, on Sparc Stamp-it performs significantly
worse than the other schemes. But with an increased workload this performance
difference completely vanishes. The reason for this is the higher overhead in
Stamp-it's enter and leave functions. With only a single element and a low
workload, this overhead dominates the total work each thread is handling. By
increasing the workload, this overhead becomes much less relevant, while at the
same time efficient reclamation of the removed elements becomes more important.
So in the scenarios with higher workload Stamp-it shows much better performance.
Obviously, an increased number of elements also reduces the relevance of this
overhead. The configuration with 32 threads and 25 elements even shows inversed
results (see Figure~\ref{fig:workload-25-elem-32-threads-no-LFRC}); in this
configuration Stamp-it clearly outperforms all the other schemes on Sparc.
Interestingly, the other architectures are largely unaffected and show no such
bias.

\subsection{Scalability with traversal length}

\begin{figure*}[!b]
  \centering
  \includegraphics[width=\textwidth]{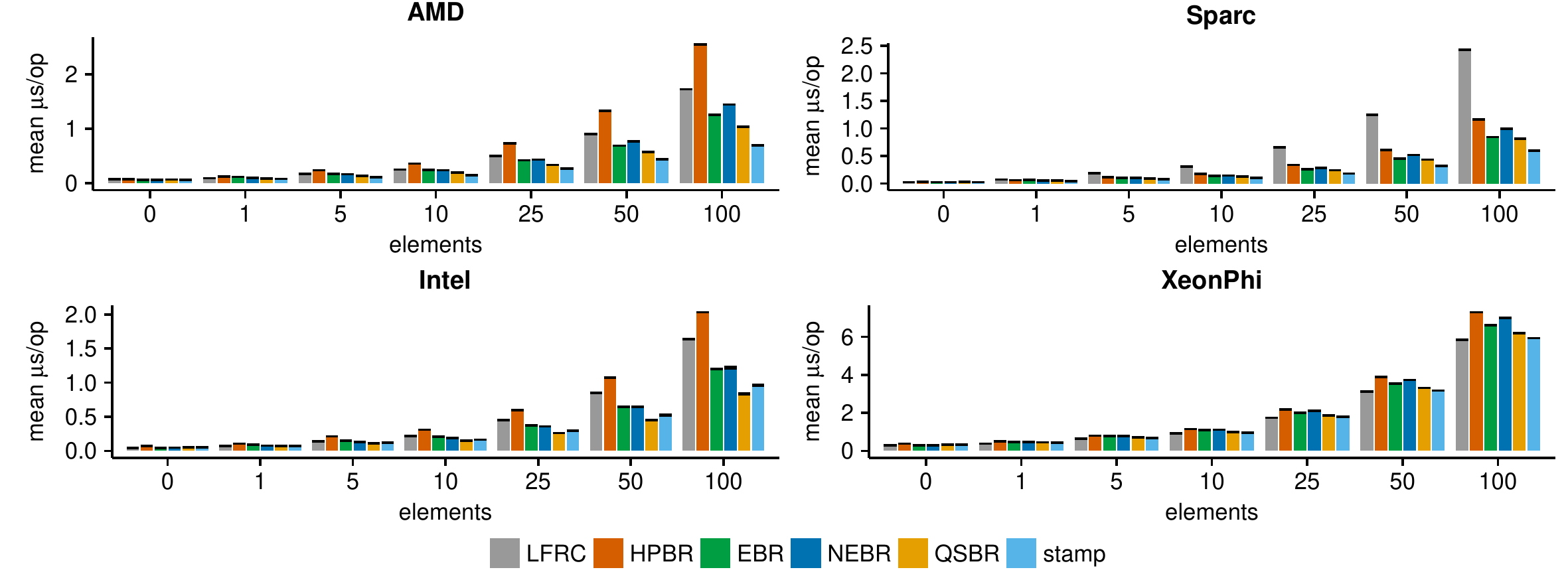}
  \caption{Effect of varying traversal length on a read-only lock-free list with one thread.}
  \label{fig:length-1-thread-0.0-workload}
\end{figure*}

\begin{figure*}
  \centering
  \includegraphics[width=\textwidth]{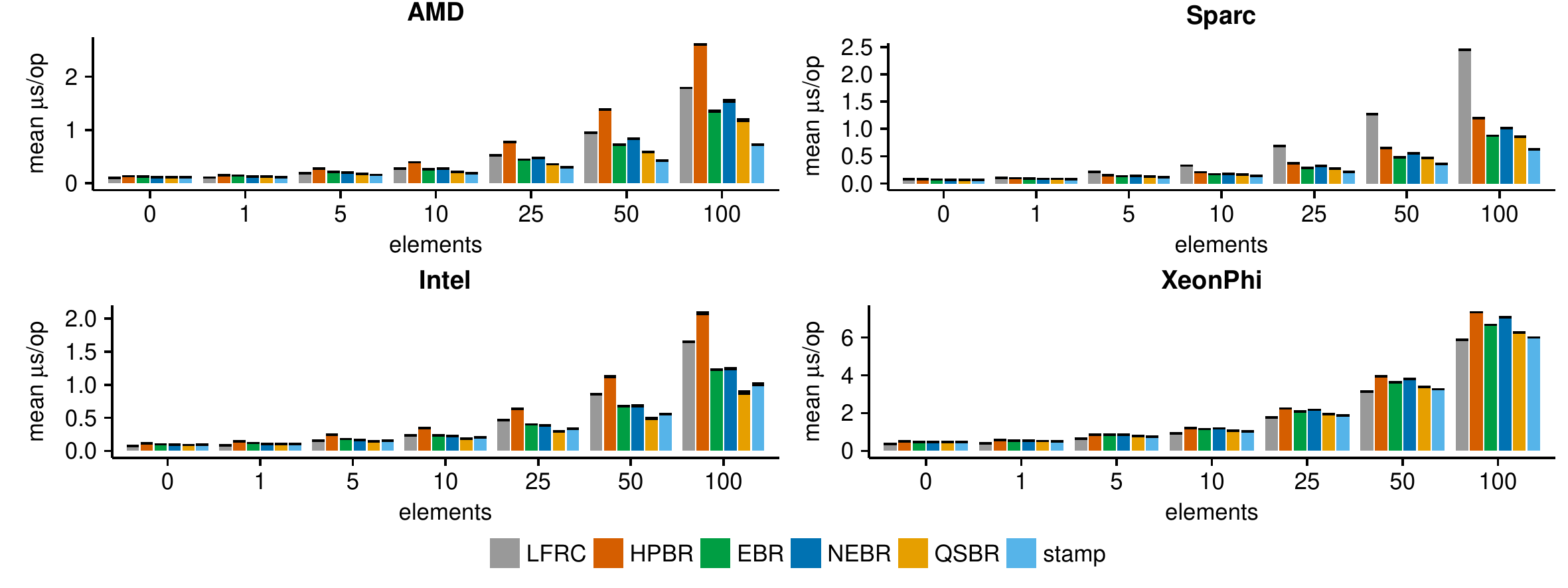}
  \caption{Effect of varying traversal length on a lock-free list with one thread and a workload of 50\%.}
  \label{fig:length-1-thread-0.5-workload}
\end{figure*}

\begin{figure*}
  \centering
  \includegraphics[width=\textwidth]{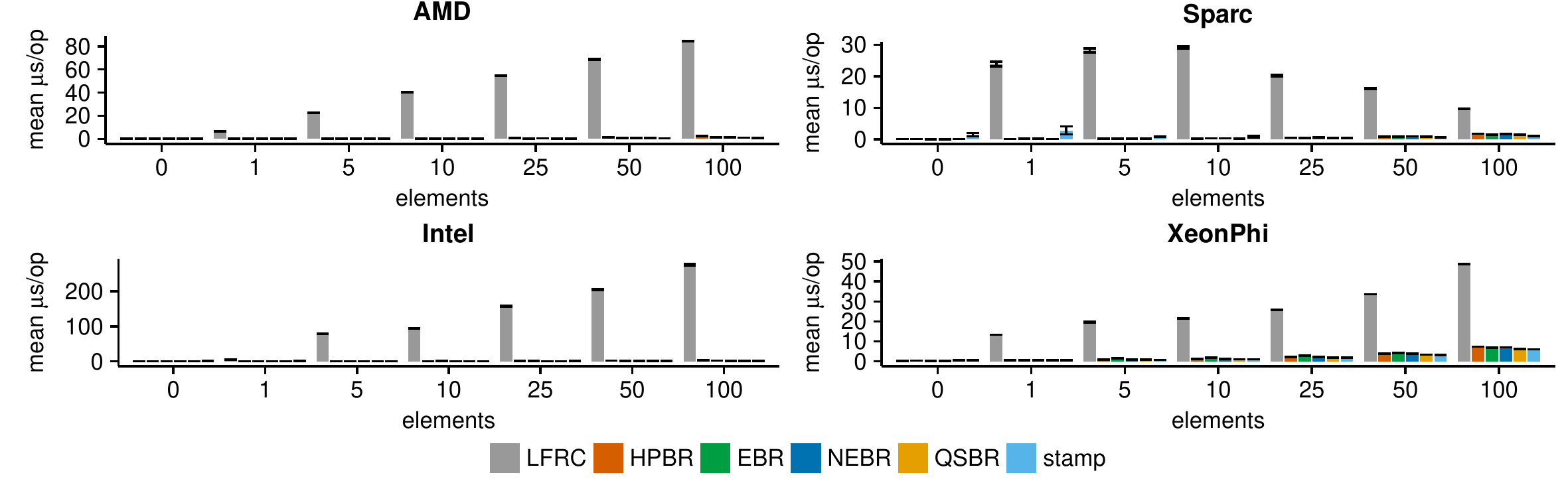}
  \caption{Effect of varying traversal length on a read-only lock-free list with 32 threads.}
  \label{fig:length-32-thread-0.0-workload}
\end{figure*}

\begin{figure*}
  \centering
  \includegraphics[width=\textwidth]{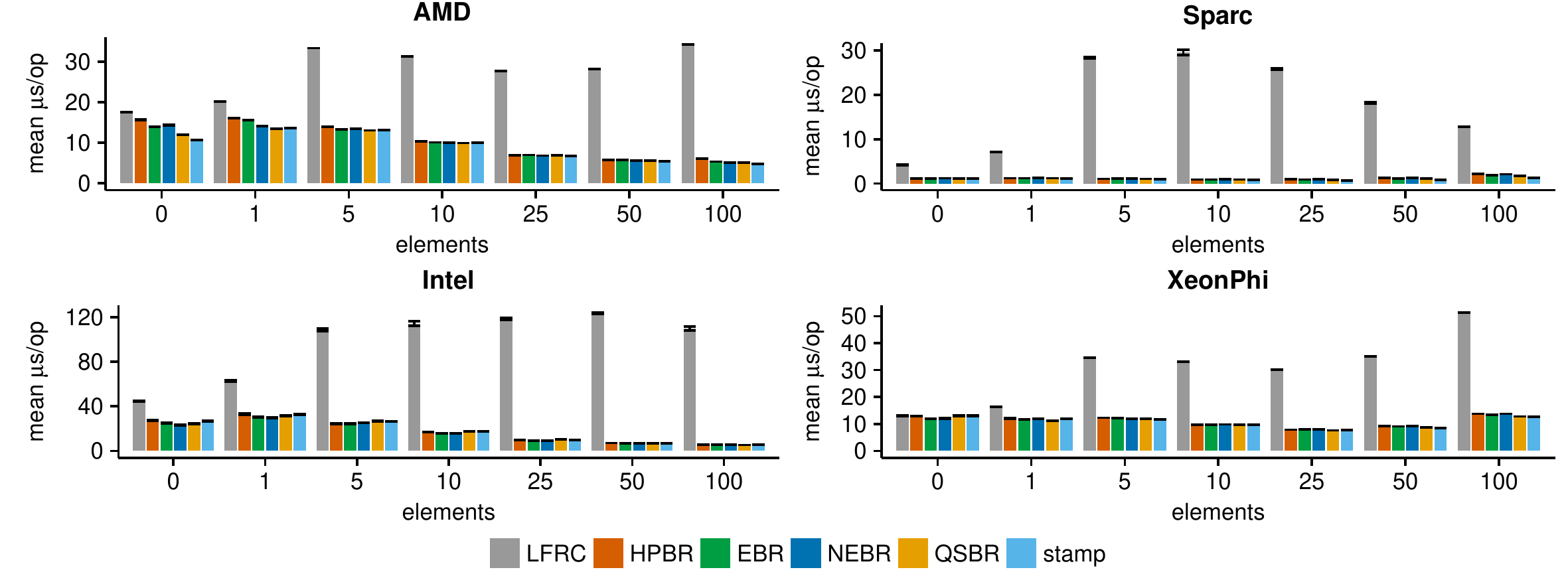}
  \caption{Effect of varying traversal length on a lock-free list with 32 threads and a workload of 50\%.}
  \label{fig:length-32-thread-0.5-workload}
\end{figure*}

\begin{figure*}[!htb]
	\centering
	\includegraphics[width=\textwidth]{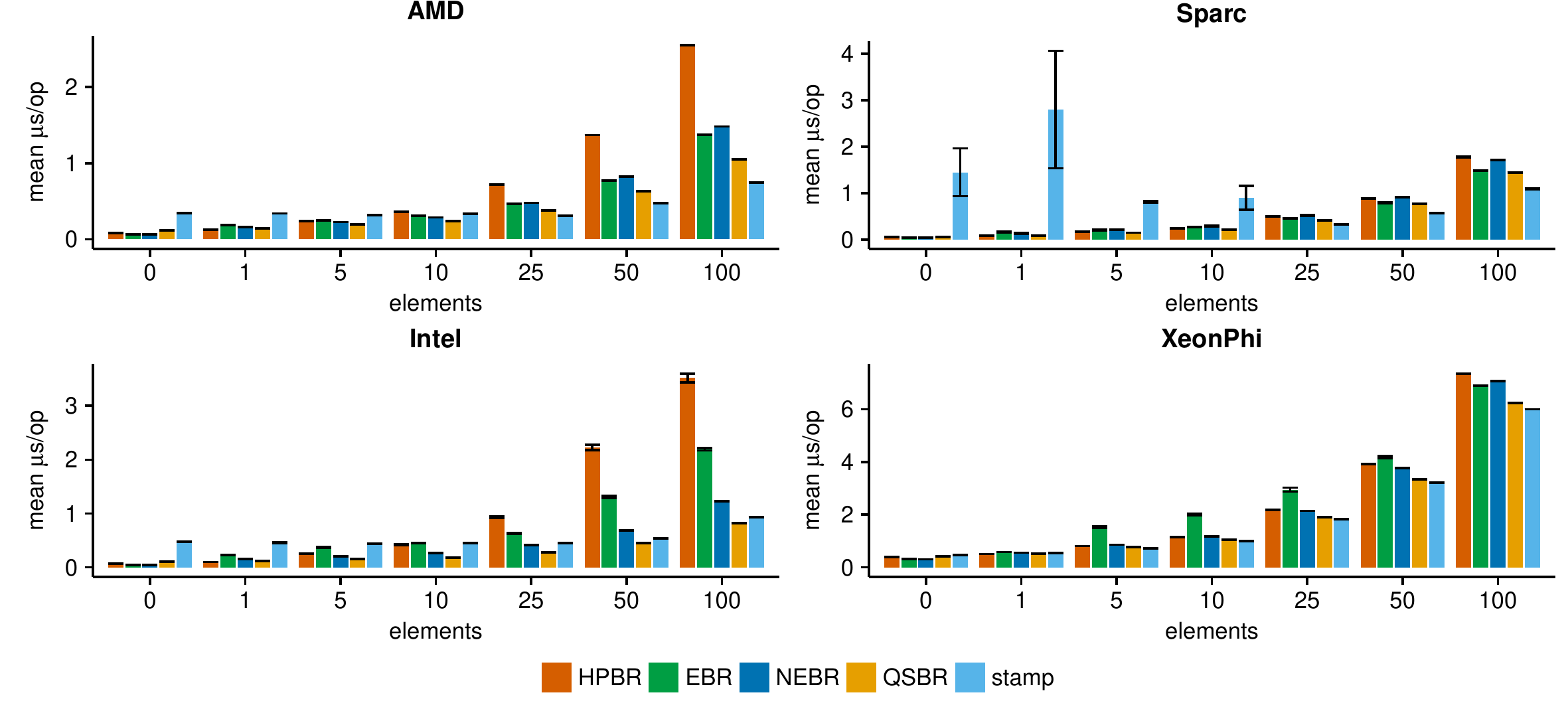}
	\caption{Effect of varying traversal length on a read-only lock-free list with 32 threads (without LFRC).}
	\label{fig:length-32-thread-0.0-workload-no-LFRC}
\end{figure*}

\begin{figure*}[!htb]
	\centering
	\includegraphics[width=\textwidth]{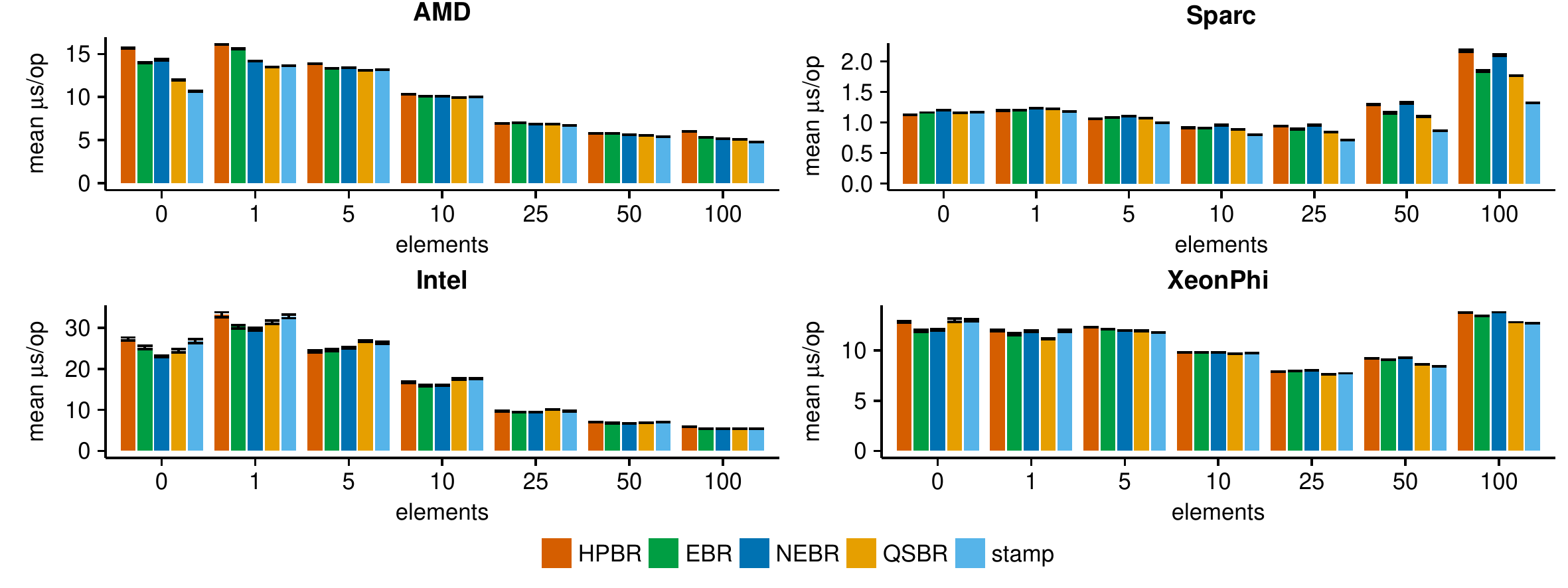}
	\caption{Effect of varying traversal length on a lock-free list with 32 threads and a workload of 50\% (without LFRC).}
	\label{fig:length-32-thread-0.5-workload-no-LFRC}
\end{figure*}

The number of elements in a list can also have an impact on how good
the different reclamation schemes perform. This analysis examines this
impact by varying the number of elements the list gets initialized with
at the start of each trial from zero to 1000. It is also run in four
different configurations, each with 30 trials and a runtime of eight
seconds:
\begin{itemize}
	\item one thread; workload of zero (\ie read-only)
		(see Figure~\ref{fig:length-1-thread-0.0-workload})
	\item one thread; workload of 50\% (see Figure~\ref{fig:length-1-thread-0.5-workload})
	\item 32 threads; workload of zero (\ie read-only)
		(see Figures~\ref{fig:length-32-thread-0.0-workload} and~\ref{fig:length-32-thread-0.0-workload-no-LFRC})
	\item 32 threads; workload of 50\% (see Figure~\ref{fig:length-32-thread-0.5-workload}
		and~\ref{fig:length-32-thread-0.5-workload-no-LFRC})
\end{itemize}

Results can be seen in Figures~\ref{fig:length-1-thread-0.0-workload},
\ref{fig:length-1-thread-0.5-workload},
\ref{fig:length-32-thread-0.0-workload}, and
\ref{fig:length-32-thread-0.5-workload}.


The single threaded results for a read-only list (see
Figure~\ref{fig:length-1-thread-0.0-workload}) and a workload of 50\%
(see Figure~\ref{fig:length-1-thread-0.5-workload}) look almost
identical. This corroborates the observations from the previous
analysis that the workload has no significant impact on the
performance of the reclamation schemes.

What can be seen from the results of the single thread configurations
is that with an increasing traversal length the performance of LFRC
and HP degrades.  This is expected since, due to their design, these
schemes have a per-element overhead. It is interesting though, that
this effect varies in intensity depending on the respective
architecture.


When looking at the results of the 32 thread configurations
(Figures~\ref{fig:length-32-thread-0.0-workload}
and~\ref{fig:length-32-thread-0.5-workload}), LFRC's runtime goes
through the roof---especially in the read-only case. Therefore
Figures~\ref{fig:length-32-thread-0.0-workload-no-LFRC}
and~\ref{fig:length-32-thread-0.5-workload-no-LFRC} show the same
results, but with LFRC excluded. From these results one can see that
in the read-only configuration the additional overhead of HP is
highly significant, but becomes negligible when looking at the results
with 50\% workload.

For all the other schemes the results suggests that the traversal
length does not have a significant impact on their respective
performance. This is not unexpected as NER, QSR and Stamp-it all
benefit from the use of a \texttt{region\_guard} to amortize overhead
over a number of operations. ER does not use the concept of
\texttt{region\_guard}s, so the number of attempts to update the
global epoch is in direct proportion to the number of created
\texttt{guard\_ptr} instances, and is thus directly linked to the
number of elements in the list. An indication of this can be seen in
the Intel results in
Figure~\ref{fig:length-32-thread-0.5-workload-no-LFRC}, but overall
this overhead is less relevant than one might have expected.

\subsection{Adding threads}

\begin{figure*}
  \centering
  \includegraphics[width=\textwidth]{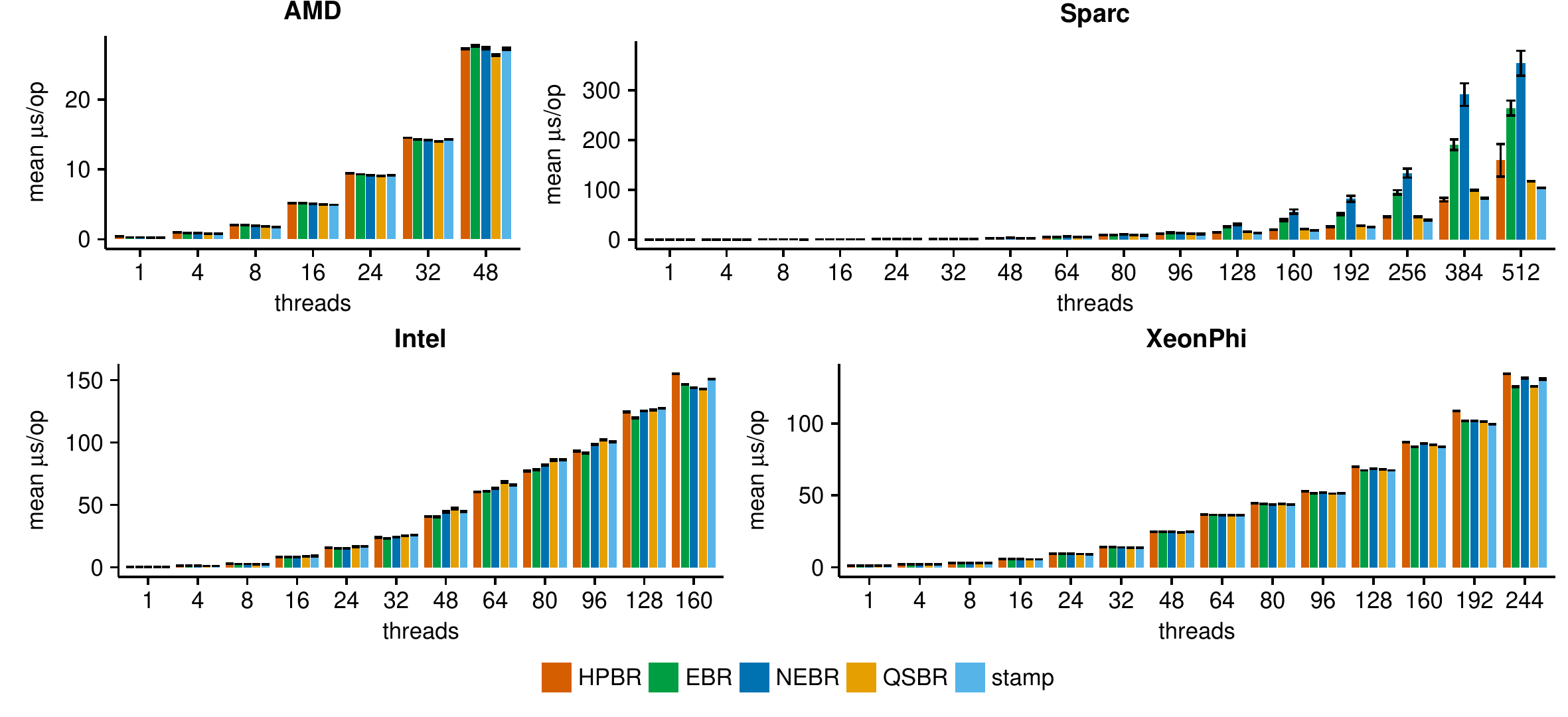}
  \caption{Performance of the List benchmark with 10 elements, a workload of 80\% and a varying number of threads (without LFRC).}
  \label{fig:threads-list-80}
\end{figure*}

Figure~\ref{fig:threads-list-80} shows the results of the List
benchmark with a workload of 80\% and a varying number of threads. These
results were excluded from the main text since they are qualitatively
similar to the results with a workload of 20\% shown in
Figure~\ref{fig:threads-list-20}.

\subsection{Reclamation efficiency}
\label{appendix:reclamantion-efficiency}

\begin{figure*}
	\centering
	\includegraphics[width=\textwidth]{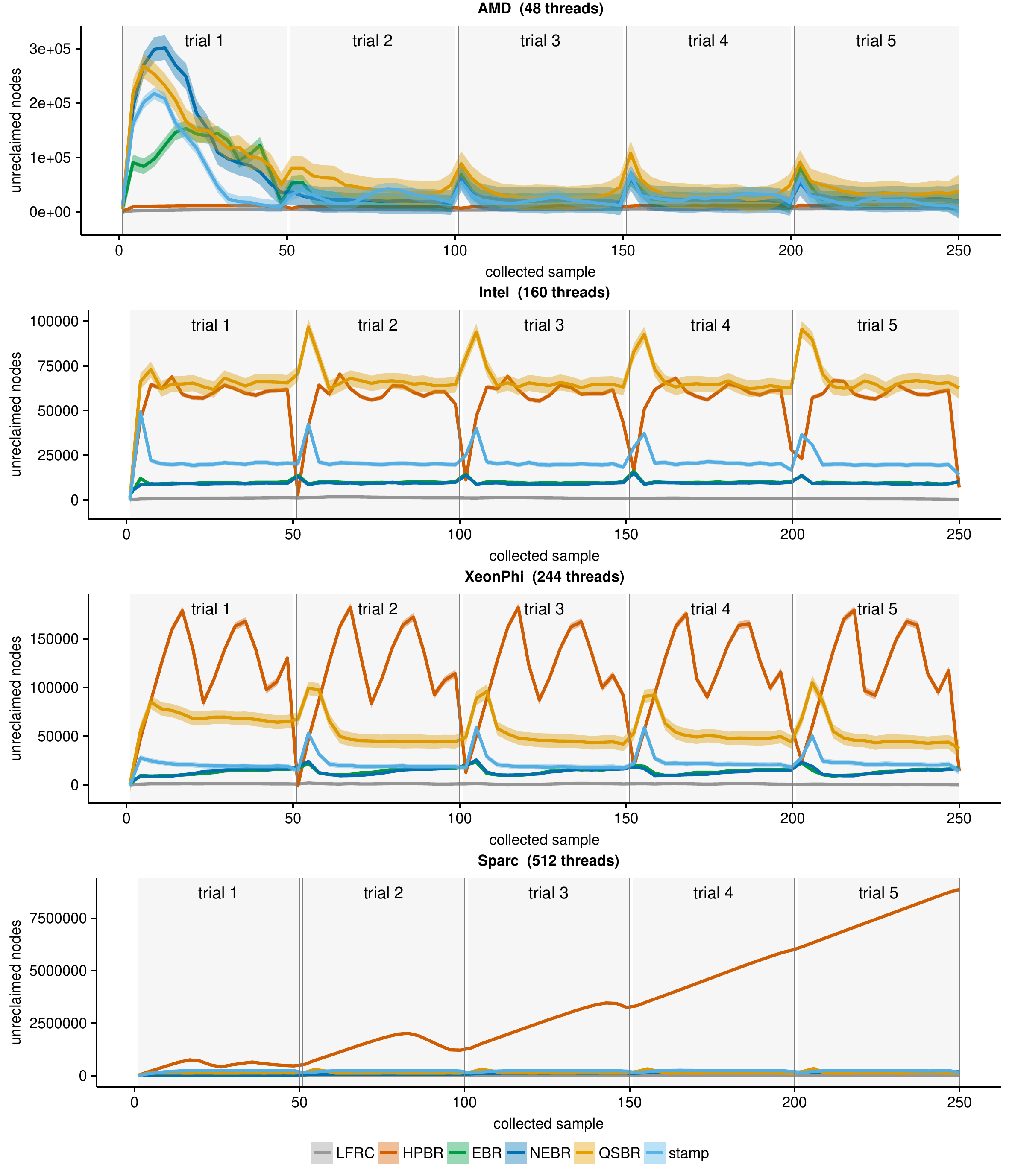}
	\caption{Number of unreclaimed of nodes over time in the Queue benchmark.}
	\label{fig:unreclaimed-nodes-queue}
\end{figure*}
\begin{figure*}
	\centering
	\includegraphics[width=\textwidth]{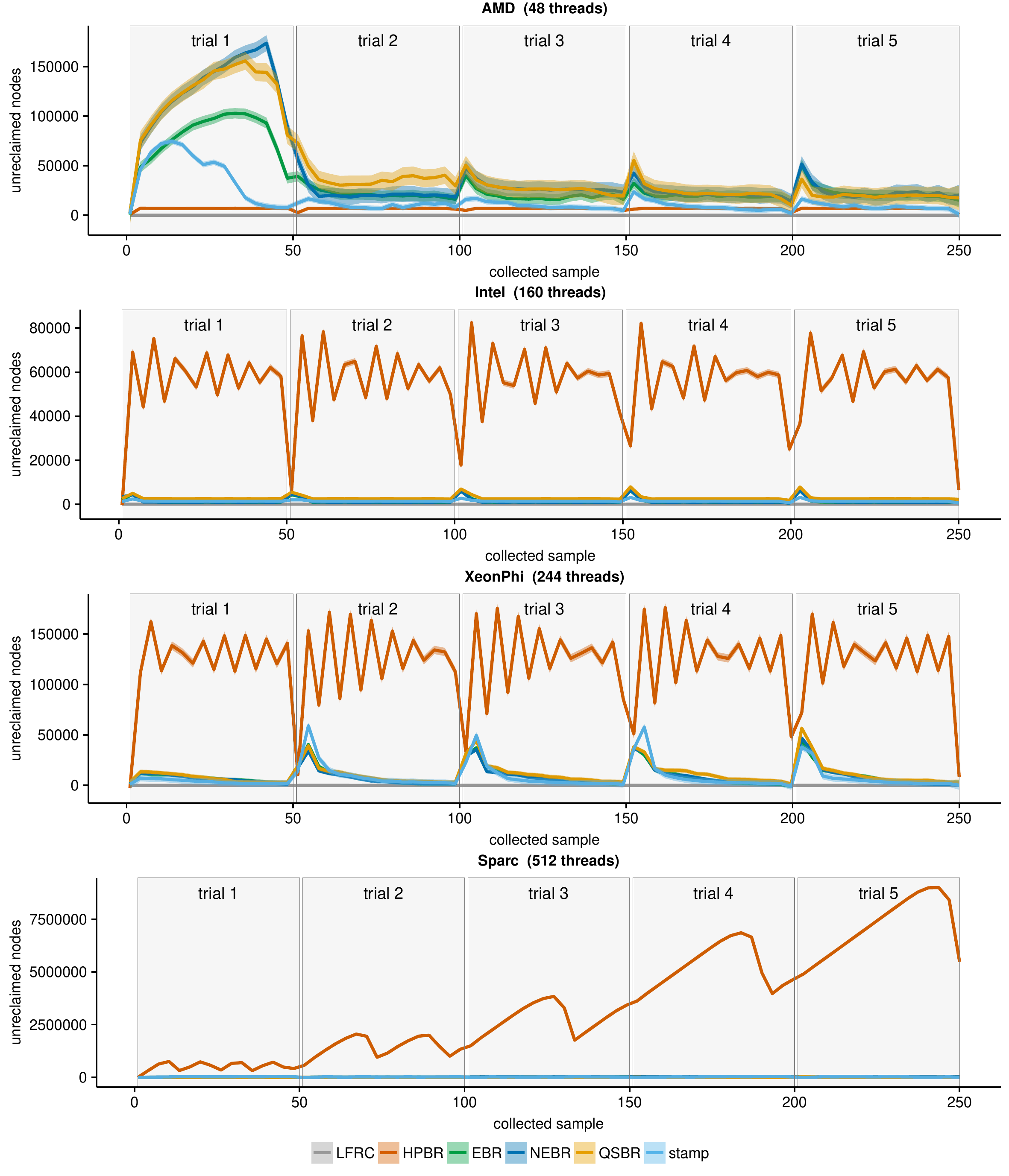}
	\caption{Number of unreclaimed nodes over time in the List benchmark with 10 elements and a workload of 20\%.}
	\label{fig:unreclaimed-nodes-list-20}
\end{figure*}
\begin{figure*}
	\centering
	\includegraphics[width=\textwidth]{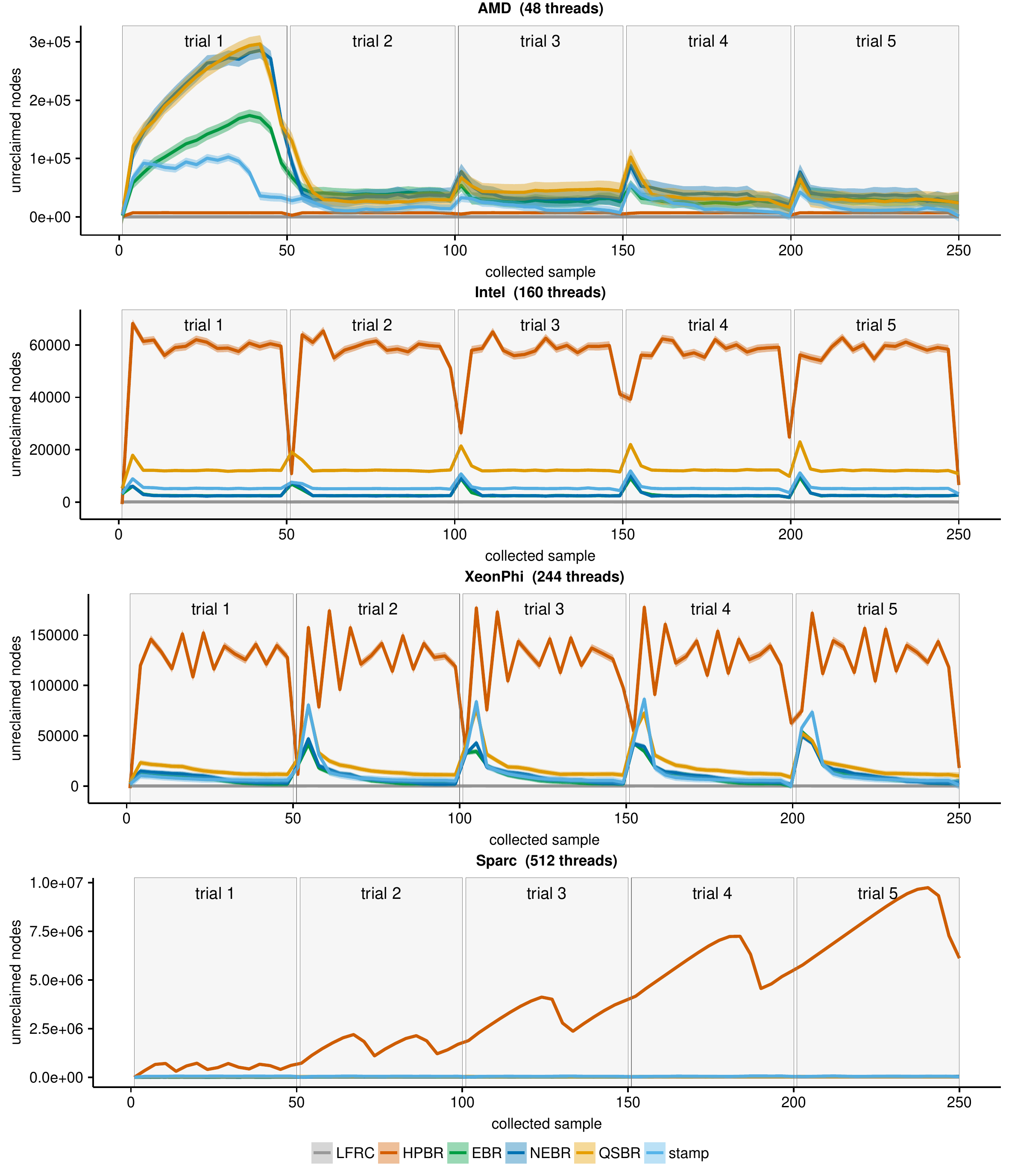}
	\caption{Number of unreclaimed nodes over time in the List benchmark with 10 elements and a workload of 80\%.}
	\label{fig:unreclaimed-nodes-list-80}
\end{figure*}
\begin{figure*}
	\centering
	\includegraphics[width=\textwidth]{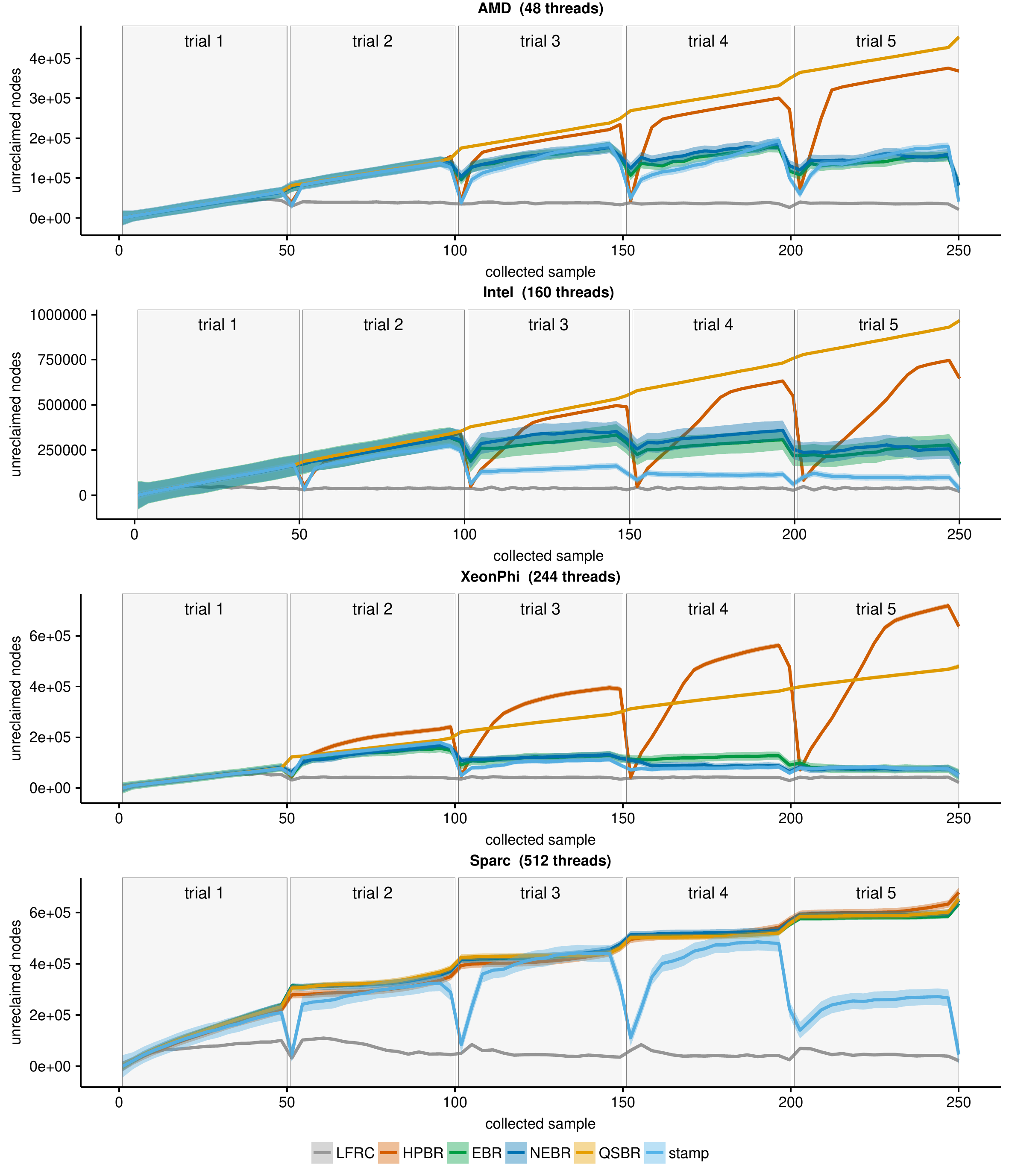}
	\caption{Number of unreclaimed nodes over time in the HashMap benchmark.}
	\label{fig:unreclaimed-nodes-hash_map}
\end{figure*}

The results are shown in Figures~\ref{fig:unreclaimed-nodes-queue},
\ref{fig:unreclaimed-nodes-list-20},
\ref{fig:unreclaimed-nodes-list-80},
and~\ref{fig:unreclaimed-nodes-hash_map}. What can be seen in all
scenarios is that HP's efficiency is inversely proportional to the
number of threads.  This is due to the fact that the threshold for the
number of unreclaimed nodes is quadratic in the number of
threads. This is the case even for the Queue benchmark
(Figure~\ref{fig:unreclaimed-nodes-queue}) and List benchmarks
(Figures~\ref{fig:unreclaimed-nodes-list-20}
and~\ref{fig:unreclaimed-nodes-list-80}), even though the number of
hazard pointers per thread is constant in these scenarios. In the
HashMap benchmark (Figure~\ref{fig:unreclaimed-nodes-hash_map}) a
dynamic number of hazard pointers is used, which makes the situation
even worse.

The implementation allows to customize the calculation of this
threshold, so for future work it might be interesting to analyse how a
different threshold would affect reclamation efficiency and what
impact this would have on the performance.

In the Queue and List benchmarks on AMD we can see a small bump in the
number of unreclaimed nodes during the first trial for all reclamation
schemes except LFRC and HP. After the first trial they all recover
and perform comparably for the rest of the benchmark. It is not
entirely clear what causes this behavior as we did not investigate
further.

Apart from this behavior and the previously described issue of HP
with a large number of threads, the results for the Queue and List
benchmarks are not too surprising; all schemes perform more or less
comparably. In the Queue benchmark QSR performs somewhat worse on
Intel and XeonPhi, but this is not unexpected as QSR is less well
suited for update heavy scenarios.

In the HashMap benchmark (Figure~\ref{fig:unreclaimed-nodes-hash_map})
we can see that QSR basically fails completely to reliably reclaim
nodes on all the architectures. The number of nodes is constantly
increasing and does not even go down at the end of the trials when all
threads are stopped. This is also the reason why QSR showed such bad
performance in the previous analysis in
Section~\ref{thread-scalability}.

For HP we can also see a consistent increase in the number of
unreclaimed nodes over time, even though this number sharply drops
right at the beginning of a new trial, but also increases again very
rapidly. The only exception is Sparc, where no such drop occurs and
the number of nodes is increasing all the time. The other schemes all
perform relatively good on all architectures; the exception again
being Sparc. On Sparc HP, ER, NER and QSR are all performing
equally bad. The number of unreclaimed nodes is constantly increasing
and does not even go down at the end of the trials when all threads
are stopped.  This effect is probably caused by the fact that in these
schemes every thread is responsible for reclaiming its own retired
nodes. In Stamp-it we know if there is some other thread lagging
behind, so we can add nodes to a global list and let that thread take
responsibility for reclaiming them. This allows Stamp-it to more
reliably reclaim nodes, especially at the end of each trial.

The failure to efficiently reclaim nodes increases memory pressure,
which has a direct impact on the
runtime. Figure~\ref{fig:unreclaimed-nodes-hash_map-runtime} shows the
development of the runtime over the five trials. On Sparc we can see
that the runtime of HP, ER, NER and QSR is increasing with each
trial, while LFRC and Stamp-it is decreasing. On the other
architectures, runtime is decreasing for all schemes except QSR. This
would be the expected behavior since more results can be reused once
the hash-map has been filled.

\end{document}